\documentclass[a4paper,11pt]{article}
\usepackage{jheppub} 

\usepackage[T1]{fontenc} 
\usepackage[utf8]{inputenc}
\usepackage{hyperref}
\usepackage{placeins}
\usepackage{mynotations}
\usepackage{amssymb}
\title{Four-Family ${\cal N}=1$ Supersymmetric Pati-Salam Models from Intersecting D6-Branes}

\author[a,b,1]{Tianjun Li}
\author[c,2]{Rui Sun}
\author[a,b,3]{Chi Zhang}

\affiliation[a]{CAS Key Laboratory of Theoretical Physics, Institute of Theoretical Physics,\\
		Chinese Academy of Sciences, Beijing 100190, P. R. China}
\affiliation[b]{School of Physical Sciences, University of Chinese Academy of Sciences,\\
		No.19A Yuquan Road, Beijing 100049, P. R. China}
\affiliation[c]{Korea Institute for Advanced Study,\\
		85 Hoegiro, Dongdaemun-Gu, Seoul 02455, Korea}

\emailAdd{tli@itp.ac.cn}
\emailAdd{sunrui@kias.re.kr}
\emailAdd{zhangchi2018@itp.ac.cn}

\abstract{We investigate the construction of four-family ${\cal N}=1$ supersymmetric Pati-Salam models from Type IIA $\mathbb{T}^6/{\mathbb{Z}_2\times \mathbb{Z}_2}$ orientifold with intersecting $D$6-branes. Utilizing the deterministic algorithm introduced in Ref.~\cite{heCompleteSearchSupersymmetric2021}, we obtain $274$ types of models with three rectangular tori and distinct gauge coupling relations at string scale, while $6$ types of models with two rectangular tori and one titled torus. In both cases, there exists a class of models with gauge coupling unification at string scale. In particular, for the models with two rectangular tori, one tilted torus and gauge coupling unification, the gaugino condensations are allowed, and thus supersymmetry breaking and moduli stabilization are possible for further phenomenological study.

}

\begin{document}

\maketitle

\section{Introduction}

One of the main motivations of string phenomenology is to find a unifying $\mathcal{N}=1$ supersymmetric quantum field theory, a competent framework among various extensions of the four-dimensional Standard Model (SM). D-branes play an important role in constructing interesting models at the phenomenological level, especially in Type I, Type IIA and Type IIB string theories. Chiral fermions appear at 
\begin{itemize}
    \item worldvolume singularities of D-branes~\cite{berkoozOrbifoldTypeStrings1997, shiuTeVScaleSuperstring1998, lykkenBranesGUTsSupersymmetry1999, cveticThreeFamilyType2000, aldazabalTypeIIBOrientifolds1998,aldazabalDBranesSingularitiesBottomUp2000, kleinOrientifoldsDiscreteTorsion2000}; and 
    \item intersecting loci of D-branes in the internal space \cite{berkoozBranesIntersectingAngles1996a}. 
\end{itemize}

Intersecting $D6\mhyphen$branes on  Type IIA orientifolds have been used to construct three-family non-supersymmetric models and grand unified models~\cite{blumenhagenNoncommutativeCompactificationsType2000,blumenhagenTypeStringsBFlux2001,aldazabalIntersectingBraneWorlds2001, aldazabalChiralStringCompactifications2001,ibanezGettingJustStandard2001,angelantonjTypeIStringsMagnetised2000,forsteOrientifoldsBranesAngles2001, forsteSupersymmetricTimesOrientifolds2001,blumenhagenStandardModelStable2001,cremadesIntersectingBraneModels2002,cremadesStandardModelIntersecting2002,cremadesSUSYQuiversIntersecting2002,bailinIntersectingD5braneModels2003,bailinNewStandardlikeModels2002, bailinStandardlikeModelsIntersecting2002, bailinStandardlikeModelsIntersecting2003, kokorelisDeformedIntersectingD6Brane2002,kokorelisDeformedIntersectingD6Brane2002a,kokorelisExactStandardModel2002, kokorelisExactStandardModel2004, kokorelisNewStandardModel2002}. Even though these models satisfy the Ramond-Ramond~(RR) tadpole cancellation conditions, there are Neveu-Schwarz-Neveu-Schwarz~(NS-NS) tadpoles remaining due to their non-supersymmetric nature. Moreover, the string scale is close to the Planck scale since the intersecting $D6\mhyphen$branes are not transversal in the internal space. As a result, there are large Planck scale corrections at the loop level, leading to the gauge hierarchy problem. As a remedy, a large number of supersymmetric standard-like models and grand unified models \cite{cveticChiralFourDimensionalSupersymmetric2001,cveticMoreSupersymmetricStandardlike2003,cveticSupersymmetricThreeFamily2004,cveticPhenomenologyThreeFamilyStandardlike2002,cveticThreeFamilySupersymmetricStandardlike2001a, cveticDynamicalSupersymmetryBreaking2003, blumenhagenSupersymmetricIntersectingBranes2003,honeckerChiralSupersymmetricModels2003, liQuasiSupersymmetricUnificationIntersecting2003a, cveticSupersymmetricPatiSalamModels2004, cveticD6braneSplittingType2005, chenKtheoryAnomalyFree2005, chenRealisticSupersymmetricSpectra2008, chenRealisticWorldIntersecting2008, chenRealisticYukawaTextures2008, chenStandardLikeModelBuilding2006, chenSupersymmetricFlippedSU2005, chenVariationsHiddenSector2008, chenYukawaCorrectionsFourPoint2008} have been constructed, with the gauge hierarchy problem solved. We refer to \cite{blumenhagenRealisticIntersectingDBrane2005} for a comprehensive review for such kind of non-supersymmetric and supersymmetric models.

Among these supersymmetric models,  Pati-Salam models has been a prominent road to the Standard Model, without adding any extra $\UnitaryGroup{1}$ symmetry around the electroweak level. In Refs.~\cite{cveticSupersymmetricPatiSalamModels2004}, Cveti\v{c}, Liu and one of us (TL) showed how to systematically construct $\mathcal{N}=1$ supersymmetric Pati-Salam models from intersecting $D6\mhyphen$branes on Type IIA $\Quotient{\Torus{6}}{\Integer_2\times \Integer_2}\mhyphen$orientifold. After $D\mhyphen$brane splitting and supersymmetric preserving Higgs mechanism applied, the Pati-Salam gauge symmetry $\SpecialUnitaryGroup{4}_C \times \SpecialUnitaryGroup{2}_L \times \SpecialUnitaryGroup{2}_R$ eventually breaks down to the SM gauge symmetry. Due to the supersymmetry breaking and moduli stabilization triggered by their two confining groups in the hidden sectors, these models do have realistic and phenomenological consequences, as shown in Refs.~\cite{chenRealisticSupersymmetricSpectra2008,  liRevisitingSupersymmetricPati2021,Li:2021pxo}.
Until very recent, intriguingly all possible $202752$ three-family $\mathcal{N}=1$ supersymmetric Pati-Salam models on $\Quotient{\Torus{6}}{\Integer_2 \times \Integer_2}$ have been found  and classified with $33$ type of independent models according to gauge coupling relations~\cite{heCompleteSearchSupersymmetric2021}. 

Most of these above achievements are based on three-family model buildings. In fact, the study of the SM with four-families is also worthy of attention due to flavor democracy hypothesis. In \cite{harariQuarkMassesCabibbo1978}, a democratic mass matrix model was introduced to mainly fix the mass gap problem between three families of fermions of the SM, as well as the hierarchy problem of the Yukawa couplings. Allowing three families of fermions, one gets typical SM predictions, such as a low mass of the top quark and the inequality between three neutrino masses \cite{sultansoyFlavorDemocracyParticle2007}. If one allows  four families of fermions in the democratic mass matrix model, three families of precisely massless neutrinos and a massive neutrino can be realized without the assumption on a larger hierarchy of the Yukawa couplings. Moreover, via a slight breaking of democracy, the three massless neutrinos obtain small masses.  Then the flavor problem of the SM, can be solved naturally by putting the flavor democracy hypothesis due to the important role democratic mass matrix model plays.

As a matter of fact, the SM does not make a theoretical prediction on the number of families. The only restriction for the number of SM families comes from the requirement made by the Quantum Chromodynamics (QCD). The asymptotic freedom of QCD provide an upper bound $8$ for the number of families as discussed in~\cite{grossUltravioletBehaviorNonAbelian1973}. And if we allow four families of fermions for SM, many open issues of SM can be solved in a natural way. For example, introducing a fourth massive generation to SM can alter the cross section and decay channels of the Higgs particles \cite{gronqvistParticlePhysicsModels}. When the Yukawa couplings of the fourth generation particles are large enough, these particles are natural candidates for electroweak symmetry breaking \cite{hungDynamicalElectroweakSymmetry2011}.
Motivated by constructing realistic four-family SM, in this paper, we concentrate on 
building four-family $\mathcal{N}=1$ supersymmetric models with gauge symmetry $\SpecialUnitaryGroup{4}_C\times\SpecialUnitaryGroup{2}_L \times \SpecialUnitaryGroup{2}_R$ on $\Quotient{\Torus{6}}{\Integer_2 \times \Integer_2}$ orientifold with intersecting D6-branes 
as an extension to \cite{cveticSupersymmetricPatiSalamModels2004, heCompleteSearchSupersymmetric2021}. 
We obtain four-family Pati-Salam models with gauge coupling unification at string scale 
or near string scale. In particular, there are models without any filler brane required as well.  

The paper is organized as follows. In Section~\ref{sec: basics} we will briefly review the basics of model building from intersecting $D6\mhyphen$branes on $\Quotient{\Torus{6}}{\Integer_2\times \Integer_2}$ orientifold. Constraints on $D6\mhyphen$brane configuration, such as RR tadpole cancellation condition and supersymmetric condition are also reviewed. In Section~\ref{sec: model_building}, we present the symmetry breaking mechanism for $\UnitaryGroup{4}_C \times \UnitaryGroup{2}_L \times \UnitaryGroup{2}_R$, {\it i.e.}, $D6\mhyphen$brane splitting and the Higgs mechanism. Four-family of chiral fermion condition and various symmetry relations, such as T-dualities are also discussed.  Section~\ref{sec: phenomenology} is devoted to the phenomenological features of the four-family models. 
For each class of model, we list its chiral spectrum for the open string sector. 
In Section~\ref{sec: conclusion}, we draw conclusions and briefly discuss the limitations of our work. 
The four-family Pati-Salam models are presented in the Appendix.

\section{Basics of $\Quotient{\Torus{6}}{(\Integer_2 \times \Integer_2)}\mhyphen$Orientifolds Model Construction}\label{sec: basics}

To construct realistic and four-family supersymmetric models, we recall the basics of model construction from Type IIA string theory compactified on $\Quotient{\Torus{6}}{(\Integer_2 \times \Integer_2)}\mhyphen$orientifold, with $D6\mhyphen$branes intersecting at general tilted angles, under similar settings as in \cite{cveticChiralFourDimensionalSupersymmetric2001} and \cite{cveticSupersymmetricThreeFamily2004}. 

We begin with the orientifold $\Quotient{\Torus{6}}{(\Integer_2 \times \Integer_2)}$, where $D6\mhyphen$branes can be naturally viewed as general $3\mhyphen$cycles. Using the canonical isomorphism $\Torus{6}\simeq \Torus{2}\times \Torus{2}\times \Torus{2}$, we can easily write down the coordinate chart of the orientifold. Let $z_i, i =1, 2, 3$ be the complex coordinates of the $i$th torus, respectively. Let $\theta$ and $\omega$ be the two generators of the abelian group $\Integer_2 \times \Integer_2$. In coordinate $(z_1, z_2, z_3)$, we define orientifold actions $\theta$ and $\omega$ of $\Integer_2 \times \Integer_2$ on $\Torus{6}$ by 
\begin{equation}
    \begin{split}
        \theta(z_1, z_2, z_3) &\coloneqq (-z_1, -z_2, z_3), \\
        \omega(z_1, z_2, z_3) &\coloneqq (z_1, -z_2, -z_3). 
    \end{split}
\end{equation}
In addition, we define actions $\Omega$ and $R$ of $\Integer_2\times \Integer_2$ on $\Torus{6}$, by $\Omega$ the parity-reversion on the world-sheet, and $R$ the complex conjugate of $\Torus{6}$ as a complex manifold. By the very definition of an orientifold, where the world-sheet parity and complex conjugate make no difference, there are exactly four components of the $\Quotient{\Torus{6}}{(\Integer_2 \times \Integer_2)}\mhyphen$orientifold, namely the image of $\Torus{6}$ under the action of $\Omega R$, $\Omega R \theta$, $\Omega R \omega$ and $\Omega R\theta \omega$. These components are objects bearing RR charges, and thus  $D6\mhyphen$branes are introduced to cancel their RR charges. 

Generally, $Dp\mhyphen$branes are $(p+1)\mhyphen$dimensional objects in the spacetime, where strings start from and land on. In our case, viewed from the internal space $\Quotient{\Torus{6}}{\Integer_2 \times \Integer_2}$, the $D6\mhyphen$branes are $3\mhyphen$dimensional objects. It is sufficient to identify $D6\mhyphen$branes and $3\mhyphen$cycles for physical considerations. 
The powerful Eilenberg-Zilber theorem tells us that 
\begin{equation*}
\begin{split}
    \Homology{3}{\Torus{6}}{\Integer} &\simeq \Homology{3}{\Torus{2}\times\Torus{2}\times\Torus{2}}{\Integer}\\
    &\simeq \Homology{1}{\Torus{2}}{\Integer}\times \Homology{1}{\Torus{2}}{\Integer}\times \Homology{1}{\Torus{2}}{\Integer}\\
    & \simeq \Integer^2\times\Integer^2\times\Integer^2. 
\end{split}
\end{equation*}
In the following discussion, we denote $[a_i], [b_i], i =1, 2, 3$ as generators of $\Homology{3}{\Torus{2}}{\Integer}$ for the $i$th torus respectively. Under the orientifold action $\Integer_2 \times \Integer_2$ and taking world-sheet into consideration, we find that there are only two patterns for the lattice $\Homology{1}{\Torus{2}}{\Integer}\simeq \Integer^2$: the rectangular one and the tilted one \cite{blumenhagenTypeStringsBFlux2001,cveticThreeFamilySupersymmetricStandardlike2001a, cveticSupersymmetricThreeFamily2004,chenRealisticYukawaTextures2008}. In the basis $[a_i], [b_i], i = 1, 2, 3$, the former is generated by $[a_i], [b_i]$ over $\Integer$, while the latter is generated by $[\tilde{a}_i], [b_i]$ over $\Integer$ with $[\tilde{a}_i]\coloneqq [a_i] + \frac{1}{2}[b_i]$. Under the basis $[a_1], [b_1],[a_2], [b_2], [a_3], [b_3]$, we can represent a $3\mhyphen$ cycle $[\Pi_a]$ of the orientifold $\Quotient{\Torus{6}}{(\Integer_2 \times \Integer_2)}$ by the coordinate $(n_a^1, 2^{-\beta_1}l_a^1)\times(n_a^2, 2^{-\beta_2}l_a^2)\times (n_a^3, 2^{-\beta_3}l_a^3)$, where $n_a^i, l_a^i$ are all integers, $\beta_i$ takes $0$ when the $i$th torus is rectangular and takes value $1$ for the tilted case. Under the action of $\Omega R$, a $D6\mhyphen$brane $[\Pi_a] = (n_a^1, l_a^1)\times(n_a^2, l_a^2)\times(n_a^3, l_a^3)$ becomes $[\Pi_{a'}]=(n_a^1, -l_a^1)\times(n_a^2, -l_a^2)\times(n_a^3, -l_a^3)$, with $[\Pi_{a'}]$ a short-handed notation for the image of $[\Pi_a]$.  To sum up, a $D6\mhyphen$brane $[\Pi_a]$ and its orientifold image $[\Pi_{a'}]$ take the form 

\begin{equation}
    \begin{split}
       [\Pi_a] &= (n_a^1, 2^{-\beta_1}l_a^1)\times(n_a^2, 2^{-\beta_2}l_a^2)\times(n_a^3, 2^{-\beta_3}l_a^3)\\
        [\Pi_{a'}]&=(n_a^1, -2^{-\beta_1}l_a^1)\times(n_a^2, -2^{-\beta_2}l_a^2)\times(n_a^3, -2^{-\beta_3}l_a^3).
    \end{split}
\end{equation}
Denote $[\Pi_{\Omega R}]= 2^3(1, 0)\times (1, 0) \times (1, 0)$, then we have 
\begin{equation}
    \begin{split}
        [\Pi_{\Omega R}]&= 2^3(1, 0)\times (1, 0) \times (1, 0), \\
        [\Pi_{\Omega R\omega}]&= -2^{3-\beta_2-\beta_3}(1, 0)\times (0, 1) \times (0, 1),\\
        [\Pi_{\Omega R\theta\omega}]&= -2^{3-\beta_1-\beta_3}(0, 1)\times (1, 0) \times (0, 1),\\
        [\Pi_{\Omega R\omega\theta}]&= -2^{3-\beta_1-\beta_2}(0, 1)\times (0, 1) \times (0, 1),
    \end{split}
\end{equation}
where $[\Pi_{\Omega R\omega}], [\Pi_{\Omega R\theta}], [\Pi_{\Omega R\theta\omega}]$ are the images of $[\Pi_{\omega R}]$ under the action of $\omega$, $\theta$ and $\theta \omega$ respectively. The coefficient $2^3$ comes from identifying $8$ $O6\mhyphen$branes $(\pm 1, 0)\times (\pm 1, 0) \times (\pm 1, 0)$, under the quotient of $\Omega$ and $R$ actions. The intersection number of $D6\mhyphen$branes can be easily computed as 
\begin{equation}
    \begin{split}
        I_{ab} &= [\Pi_a]\,[\Pi_b]= 2^{-k}(n_a^1\,l_b^1-n_b^1\,l_a^1)(n_a^2\,l_b^2-n_b^2\,l_a^2)(n_a^3\,l_b^3-n_b^3\,l_a^3), \\
        I_{ab'} &= [\Pi_a]\,  [\Pi_{b'}]=-2^{-k}(n_a^1\,l_b^1+n_b^1\,l_a^1)(n_a^2\,l_b^2+n_b^2\,l_a^2)(n_a^3\,l_b^3+n_b^3\,l_a^3),\\
        I_{aa'}&= [\Pi_a] \,[\Pi_{a'}] = - 2^{3-k} n_a^1\,l_a^1\,n_a^2\,l_a^2\,n_a^3\,l_a^3,
    \end{split}
\end{equation}
where $k=\beta_1 + \beta_2 + \beta_3$. If we denote  $[\Pi_{O6}]= [\Pi_{\Omega R}]+ [\Pi_{\Omega R\omega}] + [\Pi_{\Omega R\theta}] + [\Pi_{\Omega R\theta\omega}]$, we have 
\begin{equation}
    I_{a\,O6}= [\Pi_a]\, [\Pi_{O6}]= 2^{3-k}(-l_a^1\,l_a^2\,l_a^3+l_a^1\,n_a^2\,n_a^3+n_a^1\,l_a^2\,n_a^3+ n_a^1\,n_a^2\,l_a^3).
\end{equation}
The spectrum of intersecting $D6\mhyphen$branes is given in Table~\ref{tab: general_spectrum}.
\begin{table}[htbp]  
    \caption{Spectrum of intersecting $D6\mhyphen$branes.}\label{tab: general_spectrum}
    \centering
    \begin{tabular}{|c|c|}
    \hline
        Sectors &  Representations \\
        \hline \hline
        $aa$ & $\UnitaryGroup{\Quotient{N_a}{2}}$ vector multiplets\\
        & 3 adjoint chiral multiplets\\
        $ab+ba$ & $I_{ab}\,(\square_a, \overline{\square_b})$ fermions \\
        $ab'+b'a$ & $I_{ab'}\,(\square_a, \square_b)$ fermions \\
        $aa' + a'a$ & $\frac{1}{2}(I_{aa'}-\frac{1}{2}I_{a, O6}) \,\Ysymm$\, fermions \\
                 & $\frac{1}{2}(I_{aa'}+\frac{1}{2}I_{a, O6}) \,\Yasymm$ \,fermions \\
        \hline
    \end{tabular}
\end{table}

For three-family ${\cal N}=1$ supersymmetric Pati-Salam model building in Type IIA orientifolds on $\mathbb{T}^6/(\mathbb{Z}_2\times \mathbb{Z}_2)$ with intersecting D6-branes in which the $SU(4)_C\times SU(2)_L \times SU(2)_R$ gauge symmetries arise from $U(n)$ branes. To get four families of fermions, we require
\begin{equation}\label{eq: 4_generation_condition}
    \begin{split}
        I_{ab} + I_{a b'} = 4, \\
        I_{ac} = -4, \, I_{ac'}=0.
    \end{split}
\end{equation}
The conditions $I_{ab}+ I_{a b'} =4$ and $I_{ac} = -4$ ensure that there are four families of SM fermions, and the condition $I_{ac'} =0$ means that stack $a$ of $D6\mhyphen$branes  are  parallel to the orientifold image of stack $c$ of $D6\mhyphen$branes. Thus there should be open strings stretching between those two stack of $D6\mhyphen$branes. The light scalar from the NS scalar will obtain mass $Z^2_{ac'}/4\pi \alpha'$, with $Z^2_{ac'}$ the minimal squared length of the stretching string. Similarly, the light fermions from the R sector acquire the same masses \cite{aldazabalChiralStringCompactifications2001,ibanezGettingJustStandard2001,liQuasiSupersymmetricUnificationIntersecting2003a}. These light scalars and fermions form the Higgs fields needed to break the Pati-Salam gauge symmetry to the SM gauge symmetry. 

In addition, there are two main constraints for $D6\mhyphen$brane configurations and $O6\mhyphen$plane configurations, namely the {\bf\emph{RR Tadpole Cancellation Condition}} and {\bf\emph{Supersymmetry Condition}}. 
As the sources of RR charges, $D6\mhyphen$branes and $O6\mhyphen$planes should satisfy the Gauss' law, for the flux of RR fields through the compact space $\Quotient{\Torus{6}}{(\Integer_2\times\Integer_2)}$ without boundary should be conserved. This form of Gauss' law is the so-called   RR tadpole cancellation condition. To satisfy this condition,  stacks of $N_a, a=1, 2, 3$ $D6\mhyphen$branes are often needed to be introduced as the so-called filler brane, with $a$ running through three families of gauge groups. Then the RR tadpole cancellation condition reads
\begin{equation}\label{eq: RR_tadpole_condition_O6_planes}
    \sum_{a=1}^3 N_a [\Pi_a] + \sum_{a=1}^3 N_a [\Pi_{a'}]-4[\Pi_{O6}]=0, 
\end{equation}
and the coefficient $4$ before $[\Pi_{O_6}]$ comes from the $-4$ RR charges in the D6-brane charge unit.  
Other than $O6\mhyphen$planes in Equation~\eqref{eq: RR_tadpole_condition_O6_planes}, we can also introduce $D6\mhyphen$branes between the four $O6\mhyphen$planes to cancel out the total RR charges, and rewrite Equation~\eqref{eq: RR_tadpole_condition_O6_planes} free of $O6\mhyphen$planes. To do this, we  note
\begin{equation}\label{eq: A_B_C_D}
    \begin{split}
        A_a = -n_a^1 \, n_a^2  \,n_a^3, \, B_a =n_a^1  \,l_a^2  \,l_a^3,\, C_a = l_a^1  \,n_a^2  \,l_a^3, \,D_a =l_a^1 \, l_a^2 \, n_a^3;\\
        \tilde{A}_a = -l_a^1 \, l_a^2  \,l_a^3, \, \tilde{B}_a =l_a^1  \,n_a^2 \, n_a^3,\, \tilde{C}_a =n_a^1 \, l_a^2 \, n_a^3,\, \tilde{D}_a = n_a^1 \, n_a^2  \,l_a^3.
    \end{split}
\end{equation}
Therefore the tadpole cancellation conditions~\ref{eq: RR_tadpole_condition_O6_planes} can be expressed in the form 
\begin{equation}\label{eq: RR_tadpole_condition_filler_branes}
    \begin{split}
        -2^k N^{(1)} + \sum_{a =1}^3 N_a A_a = -16, \\
        -2^k N^{(2)} + \sum_{a =1}^3 N_a B_a = -16, \\
        -2^k N^{(3)} + \sum_{a =1}^3 N_a C_a = -16, \\
        -2^k N^{(4)} + \sum_{a =1}^3 N_a D_a = -16, \\
    \end{split}
\end{equation}
where $2N^{(i)}, i=1, 2, 3, 4$ are the number of filler branes between the four $O6\mhyphen$branes, as shown in Table~\ref{tab: O6_planes_configuration}. 
\begin{table}[htbp]
	\caption{Configuration of four $O6\mhyphen$planes}
	\label{tab: O6_planes_configuration}
    \centering
    \begin{tabular}{|c|c|c|}
    \hline
        Orientifold actions & $O6\mhyphen$planes & $(n^1, l^1)\times(n^2, l^2) \times (n^3, l^3)$ \\
        \hline \hline 
        $\Omega R$ & 1  & $(2^{\beta_1},  0)\times(2^{\beta_2},  0) \times (2^{\beta_3},  0)$\\
        $\Omega R \omega$ & 2  & $(2^{\beta_1},  0)\times(0, -2^{\beta_2}) \times (0, 2^{\beta_3})$\\
         $\Omega R \theta \omega$ & 3  & $(0, -2^{\beta_1})\times(2^{\beta_2}, 0) \times (0, 2^{\beta_3})$\\
         $\Omega R \theta$ & 4  & $(0, -2^{\beta_1})\times(0, 2^{\beta_2}) \times (2^{\beta_3}, 0)$\\
         \hline
    \end{tabular}
\end{table}

Furthermore, the four-dimensional $\mathcal{N}=1$ supersymmetric conditions require $\Quotient{1}{4}$ supercharges conserved under  (i)    orientation reversion of $D6\mhyphen$branes; and (ii) group action of $\Integer_2 \times \Integer_2$. 
As  shown in \cite{berkoozBranesIntersectingAngles1996a}, the four-dimensional $\mathcal{N}=1$ supersymmetry is preserved under the orientation reversion if and only if rotations of $D6\mhyphen$branes with respect to $O6\mhyphen$planes are elements of $\SpecialUnitaryGroup{3}$, while their total rotation angles equal to $0$. When the four-dimensional $\mathcal{N}=1$ supersymmetry is preserved under orientation reversion, it will be preserved under the $\Integer_2\times \Integer_2\mhyphen$action manifestly. The supersymmetric condition can be written as \cite{cveticSupersymmetricThreeFamily2004}
\begin{equation}\label{eq: SUSY_condition}
    \begin{split}
        x_A \tilde{A}_a + x_B \tilde{B}_a + x_C \tilde{C}_a + x_D \tilde{D}_a &=0\,, \\
        \Quotient{A_a}{x_A} + \Quotient{B_a}{x_B} + \Quotient{C_a}{x_C} + \Quotient{D_a}{x_D}&< 0\,, \, a = 1, 2, 3\,, 
    \end{split}
\end{equation}
with $x_A = \lambda, x_B = \lambda 2^{\beta_2 + \beta_3}/\chi_2\chi_3, x_D = \Quotient{\lambda 2^{\beta_1 + \beta_2}}{\chi_1 \chi_2}$. The positive parameter $\lambda$ is introduced to put Equation~\eqref{eq: SUSY_condition}
on an equal scaling, and $\chi_i = \Quotient{R_i^2}{R_i^1}, i = 1, 2, 3$ is the complex structure modulus parameter for the $i$th torus respectively.

\section{Gauge Symmetry Breaking via Brane Splittings}\label{sec: model_building}

To obtain SM or standard-like models via the mechanism of intersecting $D6\mhyphen$branes, there should be at least two extra $\UnitaryGroup{1}$ gauge symmetries for either supersymmetric models or non-supersymmetric models, as a result of the constraints on the quantum number of the right handed electron \cite{ibanezGettingJustStandard2001,cveticChiralFourDimensionalSupersymmetric2001, cveticMoreSupersymmetricStandardlike2003,cveticSupersymmetricThreeFamily2004}. Among these two $\UnitaryGroup{1}$ gauge symmetries, one is lepton number symmetry $\UnitaryGroup{1}_L$ and another $\UnitaryGroup{1}_{I_{3R}}$ is an analogy for right-hand weak isospin. We have the hypercharge $Q_Y$ expressed in the form 
\begin{equation}\label{eq: hypercharge}
    Q_Y = Q_{I_{3R}} + \frac{Q_B - Q_L}{2}.
\end{equation}
The baryonic charge $Q_B$ arises from $\UnitaryGroup{1}_B$, via the decomposition $\UnitaryGroup{3}_C \simeq\SpecialUnitaryGroup{3}_C \times \UnitaryGroup{1}_B$. On the other hand, since the $\UnitaryGroup{1}_{I3R}$ gauge field should be massless, the gauge group $\UnitaryGroup{1}_{I3R}$ must come from the non-abelian component of $\UnitaryGroup{2}_R$ or $\mathrm{USp}$ symmetry, otherwise the $\UnitaryGroup{1}_{I3R}$ will acquire mass from the $B\wedge F$ couplings. To get an anomaly free $\UnitaryGroup{1}_{B-L}$, the $\UnitaryGroup{1}_L$ symmetry should come from some non-abelian group for similar reasons. 
In previous studies of supersymmetric model building~\cite{cveticChiralFourDimensionalSupersymmetric2001, cveticMoreSupersymmetricStandardlike2003}, $\UnitaryGroup{1}_{I3R}$ comes from $\mathrm{USp}$ groups. These models indeed have two extra anomaly-free $\UnitaryGroup{1}$ symmetries, and have at least $8$ Higgs doublets. One could in principal break their symmetry groups down to the SM symmetry, but cannot do this without violating the D-flatness and F-flatness, thus the supersymmetry.

In this paper, as introduced in~\cite{Li:2021pxo}, we study a  generalized version of  the four-family MSSM models.  In these models the $b$ or $c$ stacks are with twice the numbers of $D6\mhyphen$branes. Then the gauge symmetries of these generalized four-family models can be broken to  the standard four-family gauge symmetries $\SpecialUnitaryGroup{4}_C \times \SpecialUnitaryGroup{2}_L \times \SpecialUnitaryGroup{2}_R$ via the Higgs mechanism. 
Taking gauge symmetries $\UnitaryGroup{4} \times \UnitaryGroup{4}_L \times \UnitaryGroup{2}_R$ as example,  we consider a $\UnitaryGroup{4}$ gauge theory with a scalar field in the adjoint representation. By choosing appropriate rotations commuting with the generators of the Lie algebra of $\SpecialUnitaryGroup{4}$, one can break $\UnitaryGroup{4}$ to $\UnitaryGroup{2}\times \UnitaryGroup{2}$, and finally to $\UnitaryGroup{2}$. We choose the rotation for $\UnitaryGroup{4}$ scalar field acting on the vacuum expectation value $\Phi_0$ as 
\begin{equation}\label{eq: U_4_rotation}
    \Phi_1 =  \begin{pmatrix}
    V_1 & 0 & 0 & 0\\
    0 & V_1 & 0 & 0\\
    0 & 0 & -V_1 & 0 \\
    0 & 0 & 0 & -V_1
    \end{pmatrix}. 
\end{equation}
We find that the $\UnitaryGroup{4}$ gauge symmetry breaks spontaneously to $\UnitaryGroup{2}\times \UnitaryGroup{2}$, as the matrix in \eqref{eq: U_4_rotation} lies in the center of the Lie algebra of $\UnitaryGroup{4}$. What left to us is to break down $\UnitaryGroup{2}\times \UnitaryGroup{2}$ to $\UnitaryGroup{2}$. For each $\UnitaryGroup{2}$ component in $\UnitaryGroup{2}\times\UnitaryGroup{2}$, the generators of its Lie algebra are the standard Pauli matrices. If we choose the rotation matrix for $\UnitaryGroup{2}\times \UnitaryGroup{2}$ as 
\begin{equation}\label{eq: U2_U2_matrix}
    \Phi_2 = \begin{pmatrix}
    0 & 0 & V_2  &0\\
    0 & 0 & 0 & V_2\\
    V_2 & 0 & 0 & 0\\
    0 & V_2 & 0 & 0
    \end{pmatrix}, 
\end{equation}
we find that $\UnitaryGroup{2}\times \UnitaryGroup{2}$ breaks down to $\UnitaryGroup{2}$. We also note that a mass of 
\begin{equation}
    m^2 = 8 g^2 V_2^2
\end{equation}
is acquired through the above process. For models with gauge symmetry $\UnitaryGroup{4}\times \UnitaryGroup{2}_L \times \UnitaryGroup{4}_R$ and $\UnitaryGroup{4}\times \UnitaryGroup{4}_L \times \UnitaryGroup{4}_R$, one can similarly breaks the symmetry down to $\UnitaryGroup{4}\times \UnitaryGroup{2}_L \times \UnitaryGroup{2}_R$ following the above procedure.  
The anomalies of the overall $\UnitaryGroup{1}$ symmetries are canceled by the generalized Green-Schwarz mechanism \cite{aldazabalChiralStringCompactifications2001,ibanezGettingJustStandard2001, cveticChiralFourDimensionalSupersymmetric2001}, while their fields get massive from the linear $B\wedge F$ couplings. 

The gauge symmetry  $\SpecialUnitaryGroup{4}_C \times \SpecialUnitaryGroup{2}_L \times \SpecialUnitaryGroup{2}_R$ can be further  broken down to SM gauge symmetry by $D6\mhyphen$brane splitting and Higgs mechanism. 
Firstly, one can split stack $a$ of $N_a = 8$ $D6\mhyphen$branes into stack $a_1$ of $N_{a_1} = 6$ $D6\mhyphen$branes and stack $a_2$ $N_{a_2} = 2$ $D6\mhyphen$branes. Then the $\UnitaryGroup{4}_C$ symmetry breaks down to $\UnitaryGroup{3}\times \UnitaryGroup{1}$. 
We denote the numbers of symmetric and anti-symmetric representations for $\SpecialUnitaryGroup{4}_C$ , $\SpecialUnitaryGroup{2}_L$ and $\SpecialUnitaryGroup{2}_R$ by $n^a_{\Ysymm}$ and $n^a_{\Yasymm}$,  $n^b_{\Ysymm}, n^b_{\Yasymm}$ and $n^c_{\Ysymm}, n^c_{\Yasymm}$. After splitting, the symmetric and anti-symmetric representations of $\SpecialUnitaryGroup{4}_C$ descend to symmetric representations of  $\SpecialUnitaryGroup{3}_C$ and $\UnitaryGroup{1}_{B-L}$, and anti-symmetric representations of $\SpecialUnitaryGroup{3}_C$. Note that there are $I_{a_1 a'_2}$ new fields, arising from the intersection of $a_1$ stack and $a_2$ stack of $D6\mhyphen$branes. The anomaly-free gauge symmetry is $\SpecialUnitaryGroup{3}_C \times \UnitaryGroup{1}_{B-L}$, a subgroup of $\SpecialUnitaryGroup{4}_C$. 

Similarly, the  stack $c$ of $N_c = 4$ $D6\mhyphen$branes can be broen into stack $c_1$ of $N_{a_1} = 2$ $D6\mhyphen$branes and stack $c_2$ $N_{a_2} = 2$ $D6\mhyphen$branes. Then the $\UnitaryGroup{2}_R$ symmetry breaks down to $\UnitaryGroup{1}_{I3R}$. The symmetric representations of $\SpecialUnitaryGroup{2}_R$ descend to the symmetric representations of $\UnitaryGroup{1}_{I3R}$ only. Also, there are $I_{c_1 c'_2}$ new fields, arising from the intersection of $c_1$ stack and $c_2$ stack of $D6\mhyphen$branes. The anomaly-free gauge symmetry is $\UnitaryGroup{1}_{I3R}$, a subgroup of $\SpecialUnitaryGroup{2}_R$.
 After splitting, the gauge symmetry of our model breaks down to $\SpecialUnitaryGroup{3}_C \times \SpecialUnitaryGroup{2}_L \times \UnitaryGroup{1}_{B-L}\times \UnitaryGroup{1}_{I3R}$ .

 To get just the SM gauge symmetry, we assume the minimal squared distance $Z^2_{a_2 c'_1}$ between $a_2$ stack and the orientifold image of $c_1$ stack of $D6\mhyphen$branes to be very small. Then there are $I_{a_2 c'_1}$ chiral multiplets of light fermions, arising from the open string stretching between $a_2$ stack of $D6\mhyphen$branes and the orientifold image of $c_1$ stack of $D6\mhyphen$branes. 
These particles break down $\SpecialUnitaryGroup{3}_C \times \SpecialUnitaryGroup{2}_L \times \UnitaryGroup{1}_{B-L}\times \UnitaryGroup{1}_{I3R}$ to the SM gauge symmetry, playing the same role as the right-handed neutrinos and their complex conjugates. Meanwhile, they preserve the D-flatness and F-flatness, thus the supersymmetry. In conclusion, the whole symmetry-breaking chain is 
\begin{equation}
    \begin{split}
        \SpecialUnitaryGroup{4}_C \times \SpecialUnitaryGroup{2}_L \times \SpecialUnitaryGroup{2}_R &\xrightarrow{a \to a_1 + a_2} \SpecialUnitaryGroup{3}_C \times \SpecialUnitaryGroup{2}_L \times \SpecialUnitaryGroup{2}_R \times U(1)_{B-L}\\
        & \xrightarrow{c\to c_1 + c_2} \SpecialUnitaryGroup{3}_C \times\SpecialUnitaryGroup{2}_L \times \UnitaryGroup{1}_{I3R} \times \UnitaryGroup{1}_{B-L}\\
        & \xrightarrow{\text{Higgs Mechanism}}\SpecialUnitaryGroup{3}_C \times \SpecialUnitaryGroup{2}_L \times \UnitaryGroup{1}_Y\,.
    \end{split}
\end{equation}
 
The process of dynamical supersymmetry breaking has been studied in \cite{cveticDynamicalSupersymmetryBreaking2003} for $D6\mhyphen$brane models from Type IIA orientifolds. 
The kinetic function for a stack $a$ of $D6\mhyphen$branes is of the form\cite{chenRealisticSupersymmetricSpectra2008} 
\begin{equation}\label{eq: kinetic_function}
    f_a = \frac{1}{4 \kappa_a} (n_a^1 \, n_a^2 \, n_a^3 s -\frac{n_a^1\, l_a^2 \, l_a^3 \, u^1}{2^{\beta_2 + \beta_3}}-\frac{l_a^1 \, n_a^2\,  l_a^3 \, u^2}{2^{\beta_1 + \beta_3}}-\frac{l_a^1\,  l_a^2\,  n_a^3 u^3}{2^{\beta_1 + \beta_2}})\,,
\end{equation}
where $\kappa_a$ is a constant with respect to the gauge groups, for instance $\kappa_a =1$ for $\SpecialUnitaryGroup{N_a}$.
 We use moduli parameters $s$ and $u^i, i =1, 2, 3$ in supergravity basis, which are related to four dimensional dilation parameter $\phi_4$ and complex structure moduli parameters $U^i,  i=1, 2, 3$ as following
\begin{equation}\label{eq: supergravity_basis_to_string_basis}
    \begin{split}
        \RealPart{s}= \frac{e^{-\phi_4}}{2\pi}\frac{\sqrt{\ImaginaryPart{U^1} \,  \ImaginaryPart{U^2}  \, \ImaginaryPart{U^3}  }   }{\Abs{U^1 U^2 U^3}  }\,, \\
    \RealPart{u^1} = \frac{e^{-\phi_4}}{2\pi} \sqrt{\frac{\ImaginaryPart{U^1}  }{\ImaginaryPart{U^2}\,  \ImaginaryPart{U^3}}  } \Abs{\frac{U^2 U^3}{U^1}}\,, \\
     \RealPart{u^2} = \frac{e^{-\phi_4}}{2\pi} \sqrt{\frac{\ImaginaryPart{U^2}  }{\ImaginaryPart{U^1}\,  \ImaginaryPart{U^3}}  } \Abs{\frac{U^1 U^3}{U^2}}\,, \\
      \RealPart{u^3} = \frac{e^{-\phi_4}}{2\pi} \sqrt{\frac{\ImaginaryPart{U^3}  }{\ImaginaryPart{U^1}\,  \ImaginaryPart{U^2}}  } \Abs{\frac{U^1 U^2}{U^3}}\,. 
    \end{split}
\end{equation}
In our present models, the $U^i, i = 1, 2, 3$ can be computed as in \cite{chenRealisticYukawaTextures2008}
\begin{equation}
    U^1 = \Imaginary \,\chi_1,\, U^2 = \Imaginary\, \chi_2, \, U^3 = \frac{2\, \chi_3^2 + 4\,\Imaginary\, \chi_3}{4+ \chi_3^2}.
\end{equation}
Moreover, the K\"ahler potential takes the form of
\begin{equation}\label{eq: Kaehler_potential}
    K = - \ln(S + \overline{S}) - \sum_{i=1}^3\ln(U^i + \overline{U}^i).
\end{equation}
Note that the three moduli parameter $\chi_1, \chi_2, \chi_3$ are not independent, as they can be expressed in terms of $x_A, x_B, x_C, x_D$ and the latter parameters are related by the supersymmetric condition~\eqref{eq: SUSY_condition}. Actually, one can determine $\chi_1, \chi_2, \chi_3$ up to an overall coefficient, namely an action of dilation on these parameters. So one has to stabilize this dilation to determine all the moduli parameters. Previous studies \cite{taylorDilatonGauginoCondensation1990, brusteinChallengesSuperstringCosmology1993, decarlosSupersymmetryBreakingDetermination1993} employ mechanisms like gaugino condensation to stabilize this overall coefficient, dictating that there should be at least two $\mathrm{USp}$ groups in the hidden sectors. Moreover, the one-loop beta functions \cite{cveticSupersymmetricPatiSalamModels2004} 
\begin{equation}\label{eq: beta_functions}
\begin{split}
    \beta^g_i &= - 3(\frac{N^{(i)}}{2} + 1) + 2 \Abs{I_{ai}} + \Abs{I_{bi}} + \Abs{I_{ci}} + 3(\frac{N^{(i)}}{2} - 1) \\
    & = -6 + 2\Abs{I_{ai} }+\Abs{I_{bi} } + \Abs{I_{ci} }
\end{split}
\end{equation}
for each $\mathrm{USp}(N^{(i)})$ arising from $2 N^{(i)}$ filler branes are required to be negative. However, in this paper, to include other potential mechanisms, we do not restrict ourselves only to models with at least two $\mathrm{USp}$ groups in the hidden sectors. 
The gauge coupling constant related to stack $a$ of $D6\mhyphen$branes is 
\begin{equation}
    g_a^{-2} = \Abs{\RealPart{f_a}}\,, 
\end{equation}
and the coupling constant of stack $b$ and stack $c$ of $D6\mhyphen$branes are determined in the same way. The kinetic function for $\UnitaryGroup{1}_Y$ is a linear combination of those for $\SpecialUnitaryGroup{4}_C$ and $\SpecialUnitaryGroup{2}_R$, as shown in \cite{blumenhagenTypeStringsBFlux2001,chenRealisticSupersymmetricSpectra2008}
\begin{equation}
    f_Y = \frac{3}{5}(\frac{2}{3} f_a + f_c)\,.
\end{equation}
The coupling constant $g_Y$ is determined by 
\begin{equation}
    g^{-2}_Y = \Abs{\RealPart{f_Y}}\,. 
\end{equation}
At tree-level, the gauge couplings have the relation 
\begin{equation}
    g_a^2 = \alpha g_b^2 = \beta \frac{5}{3}g_Y^2 = \gamma (\pi e^{-\phi_4})\,, 
\end{equation}
where $\alpha, \beta$ and $\gamma$ are ratios between the strong coupling and the weak coupling, and hypercharge coupling, respectively.

\subsection{T-Duality and its Variations}\label{sec: T_dualities}
In string theory, if two models are related by T-duality, these models are considered equivalent. By applying T-duality, one can tremendously simplify the process of searching inequivalent models. Before the discussion of T-duality, we first point out two obvious symmetries that relate equivalent models. 
\begin{enumerate}[(i)]\label{enum: DSEP}
    \item Two models are equivalent, if they are related by a permutation of three $\Torus{2}$; and 
    \item two $D6\mhyphen$models are equivalent if their wrapping numbers on any two $\Torus{2}$ are in opposite signs, while are the same on the third $\Torus{2}$. 
\end{enumerate}
The above two symmetries are known as the $D6\mhyphen$brane Sign Equivalent Principle (DSEP). 
Then, follow the convention of \cite{cveticSupersymmetricPatiSalamModels2004}, we introduce Type I and Type II T-dualities. 
Type I duality transformation acts on arbitrary two $\Torus{2}$, say the $j$th and $k$th $\Torus{2}$. The wrapping numbers on these tori transform as follows
\begin{equation}\label{eq: Type_I_T_duality}
    \begin{split}
        (n^j_a, l^j_a) \mapsto (- l^j_a, n^j_a)\,, \\
        (n^k_a, l^k_a) \mapsto (l^k_a, -n^k_a)\, ,
    \end{split}
\end{equation}
when Type I T-duality applies. Recall the definitions in Equation \eqref{eq: A_B_C_D}, Type I T-duality only makes an exchange between $(A_a, \tilde{A}_a)$, $(B_a, \tilde{B}_a)$, $(C_a, \tilde{C}_a)$, $(D_a,  \tilde{D}_a)$. Moreover, Type I duality transformation is often combined with the trivial two $\Torus{2}$ exchange, and we call the combination an extended Type I T-duality. 

As for Type II T-duality, it acts on all three different $\Torus{2}$. For instance, if we pick the $i$th, $j$th and $k$th $\Torus{2}$, the wrapping numbers on these tori transform as 
\begin{equation}\label{eq: Type_2_T_duality}
    \begin{split}
        (n^i_a, l^i_a) &\mapsto (-n_a^i, l_a^i)\, , \\
        (n^j_a, l^j_a) &\mapsto (l^j_a, n^j_a)\, ,\\
        (n^k_a, l^k_a) &\mapsto (l^k_a, n^k_a)\, .
    \end{split} 
\end{equation}
In \cite{cveticSupersymmetricPatiSalamModels2004}, Type II T-duality often combines with the interchange between $b$ and $c$ stacks of $D6\mhyphen$branes 
\begin{equation}\label{eq: b_c_exchange}
    b \leftrightarrow c
\end{equation}
associated to $\SpecialUnitaryGroup{2}_L$ and $\SpecialUnitaryGroup{2}_R$ gauge groups. 

If we composite Type I T-duality and DSEP, we will get a variation of Type II T-duality. Under this symmetry transformation, the wrapping numbers on all three $\Torus{2}$ change as 
\begin{equation} \label{eq: Type_2_T_duality_variation}
    \begin{split}
        (n_a^1, l_a^1) &\mapsto (n_a^1, -l_a^{1})\,, \\
        (n_a^2, l_a^2) &\mapsto (n_a^2, -l_a^2)\,, \\
        (n_a^3, l_a^3) &\mapsto (n_a^3, -l_a^3)\,, \\
        b&\leftrightarrow c\, . 
    \end{split}
\end{equation}
It's easy to see that under variation of Type II T-duality, only the signs of $\tilde{A}_a, \tilde{B}_a, \tilde{C}_a, \tilde{D}_a$ in Equation~\eqref{eq: A_B_C_D} change. 

However, it is worth mentioning that, the variation \eqref{eq: Type_2_T_duality_variation} of Type II T-duality is not an equivalence in our construction of four-family supersymmetric models, if the model is not invariant under $\SpecialUnitaryGroup{2}_L$ and $\SpecialUnitaryGroup{2}_R$ interchange. This observation makes sense at least phenomenologically. For a four-family supersymmetric model, one can obtain a new model by exchanging the $b$ and $c$ stacks of $D6\mhyphen$branes associated to the $\SpecialUnitaryGroup{2}_L$ and $\SpecialUnitaryGroup{2}_R$ groups, as the quantum numbers for $\SpecialUnitaryGroup{2}_L$ and $\SpecialUnitaryGroup{2}_R$ in the particle spectrum will exchange, so will the gauge couplings for these two groups at the string level. 
We will present with examples of this inequivalence in the next section.

\subsection{Supersymmetric 4-family Models}\label{sec: model_searching}


Employing the deterministic algorithm in \cite{heCompleteSearchSupersymmetric2021},  we  do not restrict the number of $\mathrm{USp}$ groups, and consider two cases, one without any titled torus, and the other with the third torus to be tilted, without loss of generality.  
Note that four-family models are completely different from the three-family models, thus the argument in \cite{cveticSupersymmetricPatiSalamModels2004} to exclude which torus is tilted or not cannot be directly applied to our case, due to the even number of generations. 

\subsubsection{Models without Tilted Torus}
We obtain six classes of $274$ supersymmetric four-family models with the deterministic algorithm introduced in \cite{heCompleteSearchSupersymmetric2021} with representative models presented in Appendix A. The classification is based on the gauge groups, T-equivalences and phenomenological considerations such as gauge coupling relations. 

Model~\ref{tab: model_000_1} is a class of its own, as it is the representative model without tilted torus achieving exact gauge coupling unification at the string scale. The Higgs-like particles in this model arise from the intersection at $b$ and $c'$ stacks of $D6\mhyphen$branes, while the Higgs doublets arise from the massless open string states in a $N = 2$ subsector and form vector-like pairs.

The second class of models has no $\mathrm{USp}$ group, which means the tadpole cancellation conditions are satisfied without any filler brane as a rare case. These models are represented by Model~\ref{tab: model_000_2} and Model~\ref{tab: model_000_3},  which are independent models of T-equivalence. 
These models are first four-family examples having no confining $\mathrm{USp}$ groups to achieve  approximate gauge coupling unification at the string scale.

The third class of models includes Model~\ref{tab: model_000_4} and Model~\ref{tab: model_000_5} with  negative $\beta$ function and  positive $\beta$ function respectively.

The fourth class of models includes Model~\ref{tab: model_000_6}-\ref{tab: model_000_8} with two $\mathrm{USp}$ groups. These models are independent under T-duality.

As discussed in \cite{brusteinChallengesSuperstringCosmology1993} there are at least two confining USp groups needed, with negative $\beta$ function, and thus allow for gaugino condensations, these models would need alternative  mechanisms to break the supersymmetry. 
Furthermore, we can observe from the spectrum tables that Model~\ref{tab: model_000_2}, \ref{tab: model_000_4} and \ref{tab: model_000_7} 
do not have the proper Higgs doublets with quantum number $(1, ~2,~\overline{2})$ under $U(4)_C\times U(2)_L \times U(2)_R$ gauge symmetry,
from neither the $b$ and $c$ or $c'$ stacks of brane intersection nor the  massless open string states in a $N = 2$ subsector. 
Thus, we do not have the SM fermion Yukawa couplings at renormalizable level which are invariant under 
the global $U(1)_C\times U(1)_L \times U(1)_R$ symmetry in these models.

The fifth class of models no less than three $\mathrm{USp}$ groups, and includes Models~\ref{tab: model_000_6}-\ref{tab: model_000_12}. Among these models, Model~\ref{tab: model_000_11} and Model~\ref{tab: model_000_12} are related by T-dualities with $b$ and $c$ stacks of branes swapping.  To see this, we show how Model~\ref{tab: model_000_11} and Model~\ref{tab: model_000_12} are related. To begin with,   $a$ stacks of $D6\mhyphen$branes in Model~\ref{tab: model_000_11} and Model~\ref{tab: model_000_12} are related by the DSEP:
\begin{equation*}
    \begin{split}
        (-1, 0) \times (-1, 1) \times (1, 1) &\xrightarrow{\text{DSEP (i)}} (-1, 0)\times (1, 1) \times (-1, 1) \\
        &\xrightarrow{\text{DSEP (i)}} (1, 0) \times (1, 1) \times (1, -1). 
    \end{split}
\end{equation*}
And the $b$ stack of $D6\mhyphen$branes in Model~\ref{tab: model_000_11} are related to the $c$ stack of $D6\mhyphen$branes in Model~\ref{tab: model_000_12}, by Type I T-duality \eqref{eq: Type_I_T_duality}, the DSEP, and the interchange \eqref{eq: b_c_exchange} of $b$ and $c$ stacks:
\begin{equation*}
    \begin{split}
        (0, 1) \times (-1, 1) \times (-3, 1) &\xrightarrow{\text{DSEP (i)}} (0, 1)\times (-3, 1) \times (-1, 1) \\
        & \xrightarrow{\text{Type I T-duality}} (0, 1) \times (-1,-3) \times (1, 1) \\
        & \xrightarrow{\text{DSEP (ii) and $b\leftrightarrow c $}} (0, 1) \times (1, 3) \times (-1, -1).
    \end{split}
\end{equation*}
While the $c$ stack of $D6\mhyphen$branes in Model~\ref{tab: model_000_11} and the $b$ stack of $D6\mhyphen$branes in Model~\ref{tab: model_000_12} are related by Type II T-duality, DSEP and the $b\leftrightarrow c$ exchange:
\begin{equation*}
    \begin{split}
        (1, 2) \times (-1, 0)\times (-1, 1) &\xrightarrow{\text{Type II T-duality}} (-1, 2) \times (0, -1)\times (1, -1) \\
        &\xrightarrow{\text{DSEP (i)}} (-1, 2) \times (1, -1) \times (0, -1)\\
        &\xrightarrow{\text{DSEP (ii) and $b\leftrightarrow c$}}(-1, 2) \times (-1, 1) \times (0, 1)\,.
    \end{split}
\end{equation*}
Even though these two models are related by the generalized T-duality as above, they are not phenomenologically equivalent. Model~\ref{tab: model_000_11} achieves $\UnitaryGroup{4}$ and $\UnitaryGroup{2}_R$ unification, 
while Model~\ref{tab: model_000_12} has $\UnitaryGroup{4}$ and $\UnitaryGroup{2}_L$ unification due to $b$ and $c$ stacks swapping. 
The Higgs particles of Model~\ref{tab: model_000_11} come from intersection of $b$ and $c'$ stacks of $D6\mhyphen$branes. Since all the beta functions of models within this class are negative, we may break the supersymmetry and stabilize the moduli via gaugino condensations.

The sixth class of models are with large wrapping numbers $5, 6, 7, 8, 9, 10, 11, 13, 15, 17$, represented by Models~\ref{tab: model_000_13}-\ref{tab: model_000_22} which did not appear in the former search. Three-family models with large wrapping number $5$ have been found in \cite{liRevisitingSupersymmetricPati2021}, but it's the first time to find four-family models with wrapping numbers at this scale.

\subsubsection{Models with One Tilted Torus}

Employing the deterministic algorithm, we obtain in total $6$ types of gauge coupling relations with represented models presented in Appendix B.

Model~\ref{tab: model_001_1} is a class of its own, as it is the only type of  models with one tilted torus achieving exact gauge coupling unification at the string scale. The Higgs particles in this model arise from the massless open string states in a $N = 2$ subsector and form vector-like pairs because the $b$ stack branes for these models are parallel to $c$ stack brane images on  the third two-tori.
Since all the beta functions are negative in this model, we can break the supersymmetry and stabilize the moduli via gaugino condensations. 

The second class of models includes Model~\ref{tab: model_001_2} and Model~\ref{tab: model_001_3}, and has no $\mathrm{USp}$ group. These models are related by type II T-duality. More specifically, we will below show how they are 
related by T-dualities explicitly. It's easy to find that the $a$ stacks of both models are related by DSEP and  Type II T-duality 
\begin{equation*}
\begin{split}
    (1, -1)\times (-1, 0) \times (-1, -1) & \xrightarrow{\text{DSEP}} (-1, 0) \times (1, -1) \times (-1, -1)\\
    & \xrightarrow{\text{Type II T-duality}} (1, 0) \times (-1, 1) \times (-1, -1)\,.
\end{split}
\end{equation*}
Note that the $b$ stack of wrapping numbers of Model~\ref{tab: model_001_3} are obtained by applying the variation of Type II T-duality and DSEP  on the $b$ stack of wrapping numbers of Model~\ref{tab: model_001_2}: 
\begin{equation*}
    \begin{split}
        (0, 1) \times (-1, -2) \times (1, 1) & \xrightarrow{\text{DSEP}} (-1, -2) \times (0, 1) \times (1, 1) \\
        & \xrightarrow{\text{variation of Type II T-duality}} (-1, 2) \times (0, -1) \times (1, -1) \\
        & \xrightarrow{\text{DSEP}} (1, -2) \times (0, -1) \times (-1, 1)\,.\\
    \end{split}
\end{equation*}
The $c$ stacks of Model~\ref{tab: model_001_2} and Model~\ref{tab: model_001_3} are related by applying DSEP twice: 
\begin{equation*}
    \begin{split}
        (1, 0)\times (1, -2) \times (1, 1) & \xrightarrow{\text{DSEP}} (1, -2) \times (1, 0) \times (1, 1) \\
        & \xrightarrow{\text{DSEP}} (1, -2)\times (-1, 0) \times (-1, -1)\,.\\
    \end{split}
\end{equation*}

Models~\ref{tab: model_001_4}-\ref{tab: model_001_7} with one $\mathrm{USp}$ group in the hidden sector are related by T-dualities in a similar way. 
The Higgs particles in Model~\ref{tab: model_001_2} again come from the massless open string states in a $N = 2$ subsector and form vector-like pairs because the $b$ stack branes for these models are parallel to $c$ stack brane on  the third two-tori.
Since there are no $\mathrm{USp}$ groups in the hidden sectors, gaugino condensations do not work in this case. One needs to stabilize the modulus and break the supersymmetry via different mechanism. 
For Model~\ref{tab: model_001_4}, there are four Higgs doublets arising from $N=2$ subsectors due to the parallel of $b$ stack branes and $c$ stack brane on the third two-tori.
In addition, there are models \ref{tb:model3} and \ref{tb:model6} with distinct gauge coupling relations at string scale.

\section{Phenomenological Analysis}\label{sec: phenomenology}
\subsection{Models without Tilted Torus}

We begin with Model~\ref{tab: model_000_1}. The gauge group of Model~\ref{tab: model_000_1} is $\UnitaryGroup{4}\times \UnitaryGroup{2}_L \times \UnitaryGroup{2}_R \times \mathrm{USp}(2)\times \mathrm{USp}(4)$. 
We tabulate the full spectrum of chiral particles of Model~\ref{tab: model_000_1} in Table~\ref{tab: spectrum_model_7}. 
Interestingly, both Model~\ref{tab: model_000_2} and Model~\ref{tab: model_000_3}  do not have any $\mathrm{USp}$ group, and then
they do not have any  exotic particles charged under $\mathrm{USp}$ groups as well.
We tabulate the full spectrum of chiral particles of Model~\ref{tab: model_000_2} below as a representative for this class. 
\begin{table}[htbp]\scriptsize
	\centering
	\caption{Spectrum of chiral particles of Model~\ref{tab: model_000_1}}
	\begin{tabular}{|c||c||c|c|c|c||c|c|}
		\hline
		Model \ref{tab: model_000_1} & $\SpecialUnitaryGroup{4}_C \times \SpecialUnitaryGroup{2}_L \times \SpecialUnitaryGroup{2}_R \times \mathrm{USp}(2)\times \mathrm{USp}(4)$  & $Q_{4C}$& $Q_{2L}$ & $Q_{2R}$ & $Q_{em}$ & $B-L$ & Field \\
		\hline \hline
		$ab$ & $4\times (4, \overline{2}, 1, 1, 1)$ & $1$ & $-1$ & $0$ & $-\frac{1}{3}, \frac{2}{3}, -1, 0$ & $\frac{1}{3}, -1$& $Q_L, L_L$\\
		$ac$ & $4 \times (\overline{4}, 1, 2, 1, 1)$ & $-1$ &$0$ & $1$ & $\frac{1}{3}, -\frac{2}{3}, 1, 0$ & $-\frac{1}{3}, 1$ & $Q_R, L_R$ \\
		$bc'$ & $8 \times (1, \overline{2}, \overline{2}, 1, 1)$ & $0$ &$-1$ & $-1$ & $0, \pm 1$ & $0$ & $H'$ \\
		
		$b1$ & $4 \times (1, \overline{2}, 1, 2, 1)$ & $0$ &$-1$ & $0$ & $\mp \frac{1}{2}$ & $0$ &  \\
		$a4$ & $2 \times (4, 1, 1, 1, \overline{4})$ & $1$ &$0$ & $0$ & $\frac{1}{6}, -\frac{1}{2}$ & $\frac{1}{3}, -1$ &  \\
		$c4$ & $2 \times (1, 1, \overline{2}, 1, 4)$ & $0$ &$0$ & $-1$ & $\pm \frac{1}{2}$ & $0$ &  \\
		$a_{\overline{\Ysymm}}$ & $1 \times (\overline{10}, 1, 1, 1, 1)$ & $-2$ & $0$& 0& $-\frac{1}{3}, 1$ & $-\frac{2}{3}, 2$ & \\   
		$a_{\Yasymm}$ & $1 \times (6, 1, 1, 1, 1)$ & $2$ & $0$& 0& $\frac{1}{3}, -\frac{1}{3}, -1$ & $\frac{2}{3}, -2$ & \\   
		$b_{\Ysymm}$ & $3 \times (1, 3, 1, 1, 1)$ & $0$ & $2$& 0& $0, \pm 1$ & $0$ & \\
		$b_{\overline{\Yasymm}}$ & $3\times(1, 1, 1, 1, 1)$ & $0$ & $2$ & $0$ &$0$ & $0$ & \\
		$c_{\Ysymm}$ & $1\times (1, 1, 3, 1, 1)$ & $0$ & $0$ & $2$ & $0, \pm 1$ & 0 & \\
		$c_{\overline{\Yasymm}}$ & $1 \times (1, 1, 1, 1, 1)$ & $0$ & $0$& $2$ & $0$ & $0$ & \\
		\hline\hline 
		$bc$ & $2 \times (1, \overline{2},2, 1, 1)$ & $0$ &$-1$ & $1$ & $0, \pm 1$ & $0$ & $H^i_u, H^i_d$\\
		& $2 \times (1, 2,\overline{2}, 1, 1)$ & $0$ &$1$ & $-1$ &  &  & \\
		\hline
	\end{tabular}
	\label{tab: spectrum_model_7}
\end{table}
\begin{table}[htbp]\scriptsize
    \centering
    \caption{Spectrum of chiral particles of Model~\ref{tab: model_000_2}}
    \begin{tabular}{|c||c||c|c|c|c||c|c|}
    \hline
        Model \ref{tab: model_000_2} & $\SpecialUnitaryGroup{4}_C \times \SpecialUnitaryGroup{2}_L \times \SpecialUnitaryGroup{2}_R$ & $Q_{4C}$& $Q_{2L}$ & $Q_{2R}$ & $Q_{em}$ & $B-L$ & Field \\
        \hline \hline
        $ab$ & $8\times (4, \overline{2}, 1)$ & $1$ & $-1$ & $0$ & $-\frac{1}{3}, \frac{2}{3}, -1, 0$ & $\frac{1}{3}, -1$& $Q_L, L_L$\\
        $ab'$ & $4 \times (\overline{4}, \overline{2}, 1)$ & $-1$ &$-1$ & $0$ & $\frac{1}{3}, -\frac{2}{3}, 1, 0$ & $-\frac{1}{3}, 1$ & $Q_L, L_L$ \\
        $ac$ & $4 \times (\overline{4}, 1, 2)$ & $-1$ &$0$ & $1$ & $\frac{1}{3}, -\frac{2}{3}, 1, 0$ & $-\frac{1}{3}, 1$ & $Q_R, L_R$ \\
        $bc$ & $16 \times (1, \overline{2}, 2)$ & $0$ &$-1$ & $1$ & $0 \pm 1$ & $0$ & $H'$ \\
        $b_{\Ysymm}$ & $5 \times (1, 3, 1)$ & $0$ & $2$& 0& $0, \pm 1$ & $0$ & \\
        $b_{\overline{\Yasymm}}$ & $5\times(1, 1, 1)$ & $0$ & $2$ & $0$ &$0$ & $0$ & \\
         $c_{\overline{\Ysymm}}$ & $7\times (1, 1, \overline{3})$ & $0$ & $0$ & $-2$ & $0, \pm 1$ & 0 & \\
         $c_{\overline{\Yasymm}}$ & $9 \times (1, 1, 1)$ & $0$ & $0$& $2$ & $0, \pm 1$ & $0$ & \\
        \hline\hline 
         $bc'$ & $8\times (1, \overline{2}, \overline{2})$ & $0$ &$-1$ & $-1$ & $0, \pm 1$ & $0$ & $H'$\\
              & $8 \times (1, 2, 2)$ & $0$ &$1$ & $1$ &  &  & \\
        \hline
    \end{tabular}
    \label{tab: spectrum_model_000_2}
\end{table}

Model~\ref{tab: model_000_4} and Model~\ref{tab: model_000_5} are constructed with one $\mathrm{USp}$ group. We tabulate the full spectrum of chiral particles of Model~\ref{tab: model_000_4} in Table~\ref{tab: spectrum_model_000_4} as a representative of this class of models. 
\begin{table}[htbp]\scriptsize
    \centering
    \caption{{Spectrum of chiral particles of Model~\ref{tab: model_000_4}}}
    \begin{tabular}{|c||c||c|c|c|c||c|c|}
    \hline
        Model \ref{tab: model_000_4} & $\SpecialUnitaryGroup{4}_C \times \SpecialUnitaryGroup{2}_L \times \SpecialUnitaryGroup{2}_R \times\mathrm{USp}(2)$  & $Q_{4C}$& $Q_{2L}$ & $Q_{2R}$ & $Q_{em}$ & $B-L$ & Field \\
        \hline \hline
        $ab$ & $6\times (4, \overline{2}, 1, 1)$ & $1$ & $-1$ & $0$ & $-\frac{1}{3}, \frac{2}{3}, -1, 0$ & $\frac{1}{3}, -1$& $Q_L, L_L$\\
        $ab'$ & $2 \times (\overline{4}, \overline{2}, 1, 1)$ & $-1$ &$-1$ & $0$ & $\frac{1}{3}, -\frac{2}{3}, 1, 0$ & $-\frac{1}{3}, 1$ & $Q_L, L_L$ \\
        $ac$ & $4 \times (\overline{4}, 1, 2, 1)$ & $-1$ &$0$ & $1$ & $\frac{1}{3}, -\frac{2}{3}, 1, 0$ & $-\frac{1}{3}, 1$ & $Q_R, L_R$ \\
        $bc$ & $8 \times (1, \overline{2}, 2, 1)$ & $0$ &$-1$ & $1$ & $0, \pm 1$ & $0$ & $H'$ \\
        
        $c2$ & $4 \times (1, 1, 2, \overline{2})$ & $0$ &$0$ & $1$ & $\pm \frac{1}{2}$ & $0$ &  \\
        $b_{\Ysymm}$ & $3 \times (1, 3, 1, 1)$ & $0$ & $2$& 0& $0, \pm 1$ & $0$ & \\
        $b_{\overline{\Yasymm}}$ & $3\times(1, 1, 1, 1)$ & $0$ & $2$ & $0$ &$0$ & $0$ & \\
         $c_{\overline{\Ysymm}}$ & $7\times (1, 1, \overline{3}, 1)$ & $0$ & $0$ & $-2$ & $0, \pm 1$ & 0 & \\
         $c_{\Yasymm}$ & $9 \times (1, 1, 1, 1)$ & $0$ & $0$& $2$ & $0$ & $0$ & \\
        \hline\hline 
         $bc'$ & $6 \times (1, \overline{2},\overline{2}, 1)$ & $0$ &$-1$ & $-1$ & $0, \pm 1$ & $0$ & $H'$\\
              & $6 \times (1, 2, 2, 1)$ & $0$ &$1$ & $1$ &  &  & \\
        \hline
    \end{tabular}
    \label{tab: spectrum_model_000_4}
\end{table}

Models~\ref{tab: model_000_6}-\ref{tab: model_000_8} are built with two $\mathrm{USp}$ groups. The gauge groups for Model~\ref{tab: model_000_6}, Model~\ref{tab: model_000_7} and Model~\ref{tab: model_000_8} are $\UnitaryGroup{4}\times \UnitaryGroup{2}_L \times \UnitaryGroup{2}_R \times \mathrm{USp}(2) \times \mathrm{USp}(4)$, $\UnitaryGroup{4}\times \UnitaryGroup{4}_L \times \UnitaryGroup{2}_R \times \mathrm{USp}(2) \times \mathrm{USp}(4)$, $\UnitaryGroup{4}\times \UnitaryGroup{2}_L \times \UnitaryGroup{2}_R \times \mathrm{USp}(4)^2 $ and $\UnitaryGroup{4}\times \UnitaryGroup{4}_L \times \UnitaryGroup{2}_R \times \mathrm{USp}(4) \times \mathrm{USp}(12)$, respectively. 
We tabulate the full spectrum of chiral particles in Model~\ref{tab: model_000_7} in Table~\ref{tab: spectrum_model_000_7}.
\begin{table}[htbp]\scriptsize
    \centering
     \caption{{Spectrum of chiral particles of Model~\ref{tab: model_000_7}}}
    \begin{tabular}{|c||c||c|c|c|c||c|c|}
    \hline
        Model \ref{tab: model_000_7} & $\SpecialUnitaryGroup{4}_C \times \SpecialUnitaryGroup{2}_L \times \SpecialUnitaryGroup{2}_R \times \mathrm{USp}(4)^2$  & $Q_{4C}$& $Q_{2L}$ & $Q_{2R}$ & $Q_{em}$ & $B-L$ & Field \\
        \hline \hline
        $ab$ & $6\times (4, \overline{2}, 1, 1, 1)$ & $1$ & $-1$ & $0$ & $-\frac{1}{3}, \frac{2}{3}, -1, 0$ & $\frac{1}{3}, -1$& $Q_L, L_L$\\
        $ab'$ & $2 \times (\overline{4}, \overline{2}, 1, 1, 1)$ & $-1$ &$-1$ & $0$ & $\frac{1}{3}, -\frac{2}{3}, 1, 0$ & $-\frac{1}{3}, 1$ & $Q_L, L_L$ \\
        $ac$ & $4 \times (\overline{4}, 1, 2, 1, 1)$ & $-1$ &$0$ & $1$ & $\frac{1}{3}, -\frac{2}{3}, 1, 0$ & $-\frac{1}{3}, 1$ & $Q_R, L_R$ \\
        $bc$ & $4 \times (1, \overline{2}, 2, 1, 1)$ & $0$ &$-1$ & $1$ & $0, \pm 1$ & $0$ & $H'$ \\
        
        $c2$ & $1 \times (1, 1, \overline{2}, 4, 1)$ & $0$ &$0$ & $-1$ & $\mp \frac{1}{2}$ & $0$ &  \\
         $a4$ & $2 \times (\overline{4}, 1, 1, 1, 4)$ & $-1$ &$0$ & $0$ & $-\frac{1}{6}, \frac{1}{2}$ & $-\frac{1}{3}, 1$ &  \\
          $b4$ & $3 \times (1, 2, 1, 1, \overline{4})$ & $0$ &$1$ & $0$ & $\pm \frac{1}{2}$ & $0$ &  \\
           $c4$ & $4 \times (1, 1, 2, 1, \overline{4})$ & $0$ &$0$ & $1$ & $\pm \frac{1}{2}$ & $0$ &  \\
        $a_{\Ysymm}$ & $1 \times (10, 1, 1, 1, 1)$ & $2$ & $0$& 0& $\frac{1}{3}, -\frac{1}{3}, -1$ & $\frac{2}{3}, -2$ & \\   
         $a_{\overline{\Yasymm}}$ & $1 \times (\overline{6}, 1, 1, 1, 1)$ & $-2$ & $0$& 0& $-\frac{1}{3}, \frac{1}{3}, 1$ & $-\frac{2}{3}, 2$ & \\   
        $b_{\Ysymm}$ & $1 \times (1, 3, 1, 1, 1)$ & $0$ & $2$& 0& $0, \pm 1$ & $0$ & \\
        $b_{\overline{\Yasymm}}$ & $1\times(1, 1, 1, 1, 1)$ & $0$ & $2$ & $0$ &$0$ & $0$ & \\
         $c_{\Ysymm}$ & $5\times (1, 1, 3, 1, 1)$ & $0$ & $0$ & $2$ & $0, \pm 1$ & 0 & \\
         $c_{\Yasymm}$ & $27 \times (1, 1, 1, 1, 1)$ & $0$ & $0$& $2$ & $0$ & $0$ & \\
        \hline\hline 
         $bc'$ & $7 \times (1, \overline{2},\overline{2}, 1, 1)$ & $0$ &$-1$ & $-1$ & $0, \pm 1$ & $0$ & $H'$\\
              & $7 \times (1, 2,2, 1, 1)$ & $0$ &$1$ & $1$ &  &  & \\
        \hline
    \end{tabular}
    \label{tab: spectrum_model_000_7}
\end{table}

Models~\ref{tab: model_000_9}-\ref{tab: model_000_12} are built with at least three $\mathrm{USp}$ groups. Their gauge groups are $\UnitaryGroup{4}\times\UnitaryGroup{2}_L \times \UnitaryGroup{2}_R \times \mathrm{USp}(2)^2 \times \mathrm{USp}(4)$, $\UnitaryGroup{4}\times\UnitaryGroup{2}_L \times \UnitaryGroup{2}_R \times \mathrm{USp}(2)^3 \times \mathrm{USp}(8)$, $\UnitaryGroup{4}\times\UnitaryGroup{2}_L \times \UnitaryGroup{2}_R \times \mathrm{USp}(2)^3 \times \mathrm{USp}(4)$ and $\UnitaryGroup{4}\times\UnitaryGroup{2}_L \times \UnitaryGroup{2}_R \times \mathrm{USp}(2)^3 \times \mathrm{USp}(4)$, respectively. Note that Model~\ref{tab: model_000_11} and Model~\ref{tab: model_000_12} are related by T-duality, but are not phenomenologically equivalent. This can be easily seen from the fact that Model~\ref{tab: model_000_11} has $\UnitaryGroup{4}$ and $\UnitaryGroup{2}_R$ gauge coupling unification at the string scale, while Model~\ref{tab: model_000_12} has $\UnitaryGroup{4}$ and $\UnitaryGroup{2}_L$ gauge coupling unification. 
We represent the full spectrum of chiral particles in Model~\ref{tab: model_000_11} in Table~\ref{tab: spectrum_model_000_11}.
 
\begin{table}[htbp]\scriptsize
    \centering
        \caption{Spectrum of chiral particles of Model~\ref{tab: model_000_11}}
    \begin{tabular}{|c||c||c|c|c|c||c|c|}
    \hline
        Model \ref{tab: model_000_11} & $\SpecialUnitaryGroup{4}_C \times \SpecialUnitaryGroup{2}_L \times \SpecialUnitaryGroup{2}_R \times\mathrm{USp}(2)^3\times \mathrm{USp}(4)$  & $Q_{4C}$& $Q_{2L}$ & $Q_{2R}$ & $Q_{em}$ & $B-L$ & Field \\
        \hline \hline
        $ab'$ & $4 \times (4, 2, 1, 1, 1, 1, 1)$ & $1$ &$1$ & $0$ & $-\frac{1}{3}, \frac{2}{3},- 1, 0$ & $\frac{1}{3}, -1$ & $Q_L, L_L$ \\
        $ac$ & $4 \times (\overline{4}, 1, 2, 1, 1, 1, 1)$ & $-1$ &$0$ & $1$ & $\frac{1}{3},-\frac{2}{3}, 1, 0$ & $-\frac{1}{3}, 1$ & $Q_R, L_R$ \\
        $bc$ & $2 \times (1, 2, \overline{2}, 1, 1, 1, 1)$ & $0$ &$1$ & $-1$ & $0, \pm 1$ & $0$ & $H'$ \\
        $bc'$ & $4 \times (1, \overline{2}, \overline{2}, 1, 1, 1, 1)$ & $0$ &$-1$ & $-1$ & $0, \pm 1$ & $0$ & $H$ \\
        $a3$ & $1 \times (\overline{4}, 1, 1, 1, 1, 2, 1)$ & $-1$ &$0$ & $0$ & $-\frac{1}{6}, \frac{1}{2}$ & $-\frac{1}{3}, 1$ &  \\
         $a4$ & $2 \times (4, 1, 1, 1, 1, 1,\overline{4})$ & $1$ &$0$ & $0$ & $\frac{1}{6}, -\frac{1}{2}$ & $\frac{1}{3}, -1$ &  \\
         $b1$ & $1 \times (1, \overline{2}, 1, 2, 1, 1, 1)$ & $0$ &$-1$ & $0$ & $\mp \frac{1}{2}$ & $0$ &  \\
           $b2$ & $3 \times (1, 2, 1, 1, \overline{2}, 1, 1)$ & $0$ &$1$ & $0$ & $\pm \frac{1}{2}$ & $0$ &  \\
          $c2$ & $2 \times (1, 1, 2, 1, \overline{2}, 1, 1)$ & $0$ &$0$ & $1$ & $\pm \frac{1}{2}$ & $0$ &  \\
           $c4$ & $1 \times (1, 1, \overline{2}, 1, 1, 1, 4)$ & $0$ &$0$ & $-1$ & $\mp\frac{1}{2}$ & $0$ &  \\
        $b_{\overline{\Ysymm}}$ & $2 \times (1, \overline{3}, 1, 1, 1, 1, 1)$ & $0$ & $-2$& 0& $0,\pm 1$ & $0$ & \\
        $b_{\Yasymm}$ & $2\times(1, 1, 1, 1, 1, 1, 1)$ & $0$ & $2$ & $0$ &$0$ & $0$ & \\
         $c_{\overline{\Ysymm}}$ & $1\times (1, 1, \overline{3}, 1, 1, 1, 1)$ & $0$ & $0$ & $-2$ & $0, \pm 1$ & 0 & \\
         $c_{\Yasymm}$ & $1 \times (1, 1, 1, 1, 1, 1, 1)$ & $0$ & $0$& $2$ & $0$ & $0$ & \\
        \hline
    \end{tabular}
    \label{tab: spectrum_model_000_11}
\end{table}

The exotic particles  charged by $\mathrm{USp}$ groups may form bound states and composite particles at some intermediate energy scale, as the strong coupling dynamics of the $\mathrm{USp}$ groups requires. The composite particles are consistent with anomaly cancellation conditions, as in the QCD case. These composite particles thus are charged only under the SM gauge symmetry \cite{cveticPhenomenologyThreeFamilyStandardlike2002}. There are essentially two kinds of neutral bound states. The first one comes from decomposing the rank $2$ anti-symmetric representation of the $\mathrm{USp}$ groups into two fundamental representations,  and then taking the pseudo inner product of the fundamental representations. The second one comes from the rank $2N$ anti-symmetric representation of $\mathrm{USp}(2N)$ for $N \geq 2$. The first bound state is similar to a meson that is the inner product of a fundamental representation and an anti-fundamental representation of $\SpecialUnitaryGroup{3}_C$ in QCD. The second bound state is a $\mathrm{USp}(2N)$ singlet, which is an analog to a baryon being a rank $3$ anti-symmetric representation of $\SpecialUnitaryGroup{3}_C$.
Models~\ref{tab: model_000_1}, \ref{tab: model_000_7} and \ref{tab: model_000_12} contain the second kind of bound states. 
Now we take Models~\ref{tab: model_000_1}, \ref{tab: model_000_7} and \ref{tab: model_000_12} as examples to show explicitly how bound states are formed.

The composite particle spectrum for Model~\ref{tab: model_000_1} is listed in Table~\ref{tab:composite_particle_spectrum_model_000_1}. The confining group is $\mathrm{USp}(4)$, with two charged intersection. The mixing of intersection $a4$ and $c4$ results in the chiral supermultiplets $(4, 1, 2, 1, 1)$.
\begin{table}[htbp]\scriptsize
    \centering
        \caption{The composite particle spectrum for Model~\ref{tab: model_000_1}}
    \begin{tabular}{|c|c||c|c|}
    \hline
       \multicolumn{2}{|c||}{Model \ref{tab: model_000_1}}  &\multicolumn{2}{|c|}{$\SpecialUnitaryGroup{4}_C \times \SpecialUnitaryGroup{2}_L \times \SpecialUnitaryGroup{2}_R \times\mathrm{USp}(2) \times \mathrm{USp}(4)$}  \\
       \hline
        Confining Force& Intersection & Exotic Particle Spectrum &Confined Particle Spectrum\\
        \hline \hline
        $\mathrm{USp}(4)_4$ & $a4$ & $2\times (4, 1, 1, 1, \overline{4})$ & $3\times(6, 1, 1, 1, 1), 3\times (10, 1, 1, 1, 1), 4\times (4, 1, 2, 1, 1)$ \\
         & $c4$ & $2\times (1,1, 2, 1, 4)$ & $3\times (1, 1, 1, 1, 1), 3\times (1, 1, 3, 1, 1)$ \\
         \hline
         $\mathrm{USp}(2)_1$ & $b1$ & $4\times (1,2, 1, 2, 1)$ & $10 \times ( 1, 1, 1, 1, 1), 10 \times( 1, 1, 3, 1, 1)$ \\
         \hline
    \end{tabular}
    \label{tab:composite_particle_spectrum_model_000_1}
\end{table}
Moreover, the confined particle spectrum for Model~\ref{tab: model_000_7} in Table~\ref{tab:composite_particle_spectrum_model_000_7}. The confining group is $\mathrm{USp}(4)$, and has three charged intersection. Besides self-confinement, it is also viable to form mixed-confinement between sections within the same confining group. The chiral supermultiplets $(\overline{4}, 2, 1, 1, 1), (1, 2, 2, 1, 1)$ and $(\overline{4}, 1, 2 ,1, 1)$ are yielded by the mixed-confinement between intersections $a4, b4$ and $c4$. 

 For Model~\ref{tab: model_000_12}, the composite particle spectrum is given in Table~\ref{tab:composite_particle_spectrum_model_000_12}. There are two confining groups $\mathrm{USp}(4)$ and $\mathrm{USp}(2)$, each with two charged intersections. The mixed-confinement between intersections $b2, c2$ yields the chiral supermultiplet $(1, 2, 2, 1, 1, 1, 1)$, while the mixed-confinement between intersections $a4, c4$ yields the chiral supermultiplet $(4, 1, 2, 1, 1, 1, 1)$. Note that when there is only one charged intersection, mixed-confinement will not be formed, thus only the tensor representations from self-confinement are left. Checking from the composite particle spectra, one finds that no new anomaly is introduced to the remaining gauge symmetry. Thus our models are free of anomalies. The above analysis for composite particles applies for all our models except Model~\ref{tab: model_000_2} and Model~\ref{tab: model_000_3} without any confining group. 

\begin{table}[htbp]\scriptsize
    \centering
      \caption{The composite particle spectrum for Model~\ref{tab: model_000_7}}
    \begin{tabular}{|c|c||c|c|}
    \hline
       \multicolumn{2}{|c||}{Model \ref{tab: model_000_7}}  &\multicolumn{2}{|c|}{$\SpecialUnitaryGroup{4}_C \times \SpecialUnitaryGroup{2}_L \times \SpecialUnitaryGroup{2}_R \times\mathrm{USp}(4)^2$}  \\
       \hline
        Confining Force& Intersection & Exotic Particle Spectrum &Confined Particle Spectrum\\
        \hline \hline
        $\mathrm{USp}(4)_4$ & $a4$ & $2\times (\overline{4}, 1, 1, 1, 4)$ & $3\times(\overline{6}, 1, 1, 1, 1), 3\times (\overline{10}, 1, 1, 1, 1), 6\times (\overline{4}, 2, 1, 1, 1)$ \\
         & $b4$ & $3\times (1, 2, 1, 1, \overline{4})$ & $6\times (1, 1, 1, 1, 1), 6\times (1, 3, 1, 1, 1), 12 \times (1, 2, 2, 1, 1)$ \\
         & $c4$ & $4\times (1, 1, 2, 1, \overline{4})$ & $10\times (1, 1, 1, 1, 1), 10\times (1, 1, 3, 1, 1), 8\times (\overline{4}, 1, 2, 1, 1)$ \\
         \hline
         $\mathrm{USp}(4)_2$ & $c2$ & $1\times (1, 1, \overline{2}, 4, 1)$ & $1\times ( 1, 1, \overline{3}, 1, 1), 1\times (1, 1, 1, 1, 1)$ \\
         \hline
    \end{tabular}
    \label{tab:composite_particle_spectrum_model_000_7}
\end{table}

\begin{table}[htbp]\scriptsize
    \centering
     \caption{The composite particle spectrum for Model~\ref{tab: model_000_12}}
    \begin{tabular}{|c|c||c|c|}
    \hline
       \multicolumn{2}{|c||}{Model \ref{tab: model_000_12}}  &\multicolumn{2}{|c|}{$\SpecialUnitaryGroup{4}_C \times \SpecialUnitaryGroup{2}_L \times \SpecialUnitaryGroup{2}_R \times \mathrm{USp}(2)^3 \times \mathrm{USp}(4)$}  \\
       \hline
        Confining Force& Intersection & Exotic Particle Spectrum &Confined Particle Spectrum\\
        \hline \hline
        $\mathrm{USp}(4)_2$ & $b2$ & $3\times (1, 2, 1, 1, \overline{4}, 1, 1)$ & $6\times(1, 1, 1, 1, 1, 1, 1), 6\times (1, 3, 1, 1, 1, 1, 1), 6 \times (1, 2, 2, 1, 1, 1, 1)$ \\
         & $c2$ & $2\times (1, 1, 2, 1, \overline{4}, 1, 1)$ & $3\times (1, 1, 1, 1, 1, 1, 1, 1), 3 \times (1, 1, 3, 1, 1, 1, 1)$ \\
         \hline
         $\mathrm{USp}(2)_4$ & $a4$ & $2\times (4, 1, 1, 1, 1, 1, \overline{2})$ & $3\times (6, 1, 1, 1, 1, 1, 1), 3\times (10, 1, 1, 1, 1, 1, 1), 2\times(4, 1, 2, 1, 1, 1, 1)$ \\
          & $c4$ & $1\times (1, 1, 2, 1, 1, 1, 2)$ & $1\times (1, 1, 1, 1, 1, 1, 1), 1\times (1, 1, 3, 1, 1, 1, 1)$ \\
         \hline
         $\mathrm{USp}(2)_1$ & $b1$ & $1\times (1, 2, 1, \overline{2}, 1, 1, 1)$ & $1\times (1, 1, 1, 1, 1, 1, 1), 1\times (1, 3, 1, 1, 1, 1, 1)$ \\
         \hline
         $\mathrm{USp}(2)_3$ & $a3$ & $1\times (4, 1, 1, 1, 1, 2, 1)$ & $1\times(10, 1, 1, 1, 1, 1, 1), 1\times(6, 1, 1, 1, 1, 1, 1)$\\
         \hline
    \end{tabular}
    \label{tab:composite_particle_spectrum_model_000_12}
\end{table}

\FloatBarrier

\subsection{Models with One Tilted Torus}

In this section, we show basic phenomenological properties of  models with one tilted torus.  Model~\ref{tab: model_001_1} represents the models with exact gauge coupling unification at the string level so far. The gauge symmetries therein are $\UnitaryGroup{4}\times \UnitaryGroup{2}_L \times \UnitaryGroup{2}_R \times \mathrm{USp}(2)^2$ with two confining groups in the hidden sector. The full spectrum of this model is shown in Table~\ref{tab: spectrum_model_7}. 
\begin{table}[htbp]\scriptsize
    \centering
     \caption{Spectrum of chiral particles of Model~\ref{tab: model_001_1}}
    \begin{tabular}{|c||c||c|c|c|c||c|c|}
    \hline
        Model~\ref{tab: model_001_1}  & $\SpecialUnitaryGroup{4}_C \times \SpecialUnitaryGroup{2}_L \times \SpecialUnitaryGroup{2}_R \times \mathrm{USp}(2)^2$ & $Q_{4C}$& $Q_{2L}$ & $Q_{2R}$ & $Q_{em}$ & $B-L$ & Field \\
        \hline \hline
        $ab'$ & $4\times (4, 2, 1,  1, 1)$ & $1$ & $1$ & $0$ & $-\frac{1}{3}, \frac{2}{3}, -1, 0$ & $\frac{1}{3}, -1$& $Q_L, L_L$\\
        $ac'$ & $4 \times (\overline{4} ,1 ,\overline{2}, 1, 1)$ & $-1$ &$0$ & $-1$ & $\frac{1}{3}, -\frac{2}{3}, 1, 0$ & $-\frac{1}{3}, 1$ & $Q_R, L_R$ \\
        $b4$ & $4 \times (1 ,2, 1, 1 ,\overline{2})$ & $0$ &$1$ & $0$ & $\pm\frac{1}{2}$ & $0$ &  \\
        $c2$ & $4 \times (1 , 1, 2, \overline{2}, 1)$ & $0$ &$0$ & $1$ & $\pm \frac{1}{2}$ & $0$ &  \\
        $b_{\overline{\Ysymm}}$ & $3 \times (1, \overline{3}, 1, 1, 1)$ & $0$ & $-2$& 0& $0, \pm 1$ & $0$ & \\
        $b_{\Yasymm}$ & $2\times(1, 1, 1, 1, 1)$ & $0$ & $2$ & $0$ &$0$ & $0$ & \\
         $c_{\Ysymm}$ & $3\times (1, 1, 3, 1, 1)$ & $0$ & $0$ & $2$ & $0, \pm 1$ & 0 & \\
         $c_{\overline{\Yasymm}}$ & $3 \times (1, 1, 1, 1, 1)$ & $0$ & $0$& $2$ & $0$ & $0$ & \\
           \hline\hline 
         $bc$ & $ 8 \times (1, 2,\overline{2}, 1, 1)$ & $0$ &$1$ & $-1$ & $0, \pm 1$ & $0$ & $H^i_u, H^i_d$\\
              & $8 \times (1, \overline{2},2, 1, 1)$ & $0$ &$-1$ & $1$ &  &  & \\
        \hline
    \end{tabular}
    \label{tab: spectrum_model_001_1}
\end{table}

Model~\ref{tab: model_001_2} and Model~\ref{tab: model_001_3} have no $\mathrm{USp}$ group. The gauge group for the two models is $\UnitaryGroup{4}_C \times \UnitaryGroup{2}_L \times \UnitaryGroup{2}_R$. 
Since there is no $\mathrm{USp}$ group, the gaugino condensation mechanism will not work. 
Thus, one need to find other mechanism for supersymmetry breaking. Also, there are no exotic particles in these two models, as exotic particles are charged under $\mathrm{USp}$ groups. We show the full chiral spectrum in the open string sectors for Model~\ref{tab: model_001_2} in Table~\ref{tab: spectrum_model_001_2}. 

\begin{table}[htbp]\scriptsize
    \centering
        \caption{Spectrum of chiral particles of Model~\ref{tab: model_001_2}}
    \begin{tabular}{|c||c||c|c|c|c||c|c|}
    \hline
        Model \ref{tab: model_001_2} & $\SpecialUnitaryGroup{4}_C \times \SpecialUnitaryGroup{4}_L \times \SpecialUnitaryGroup{4}_R$ & $Q_{4C}$& $Q_{2L}$ & $Q_{2R}$ & $Q_{em}$ & $B-L$ & Field \\
        \hline \hline
        $ab'$ & $4\times (\overline{4}, \overline{4}, 1)$ & $-1$ & $-1$ & $0$ & $\frac{1}{3}, -\frac{2}{3}, 1, 0$ & $-\frac{1}{3}, 1$& $Q_L, L_L$\\
        $ac'$ & $4 \times (4 ,1 ,4)$ & $1$ &$0$ & $1$ & $-\frac{1}{3}, \frac{2}{3}, -1, 0$ & $\frac{1}{3}, -1$ & $Q_R, L_R$ \\
        $b_{\overline{\Ysymm}}$ & $2 \times (1, \overline{10}, 1)$ & $0$ & $-2$& 0& $0, \pm 1$ & $0$ & \\
        $b_{\Yasymm}$ & $2\times(1, 6, 1)$ & $0$ & $2$ & $0$ &$0$ & $0$ & \\
         $c_{\Ysymm}$ & $2\times (1, 1, 10)$ & $0$ & $0$ & $2$ & $0, \pm 1$ & 0 & \\
         $c_{\overline{\Yasymm}}$ & $2 \times (1, 1, 6)$ & $0$ & $0$& $2$ & $0$ & $0$ & \\
          \hline\hline 
         $bc$ & $ 4\times (1, \overline{4}, 4)$ & $0$ &$-1$ & $1$ & $0, \pm 1$ & $0$ & $H^i_u, H^i_d$\\
              & $4\times (1,4, \overline{4})$ & $0$ &$1$ & $-1$ &  &  & \\
        \hline
    \end{tabular}
    \label{tab: spectrum_model_001_2}
\end{table}

Models~\ref{tab: model_001_4}, \ref{tab: model_001_5}, \ref{tab: model_001_6}, \ref{tab: model_001_7} have only one $\mathrm{USp}(2)$ group in the hidden sector. Their gauge symmetry is all $\UnitaryGroup{4}\times \UnitaryGroup{4}_L \times \UnitaryGroup{2}_R \times \mathrm{USp}(2)$. We note that Model~\ref{tab: model_001_4} and Model~\ref{tab: model_001_5} are T-dual to each other, as well as Model~\ref{tab: model_001_6} and Model~\ref{tab: model_001_7}. But they are clearly not equivalent models at phenomenological level due to $b$ and $c$ stacks of brane swapping. In Model~\ref{tab: model_001_4}, we have $\SpecialUnitaryGroup{3}_C$ and $\UnitaryGroup{1}_Y$ gauge coupling unification at the string level, while in Model~\ref{tab: model_001_5} this gauge unification get swapped to $\SpecialUnitaryGroup{3}_C$ and $\SpecialUnitaryGroup{2}_L$ gauge coupling unification at the string level. Similarly, this $b$ and $c$ stacks of brane swapping appear between Model~\ref{tab: model_001_6} and Model~\ref{tab: model_001_7} as well. The whole spectrum of chiral particles of Model~\ref{tab: model_001_4} in Table~\ref{tab: spectrum_model_001_4}. 
\begin{table}[htbp]\scriptsize
    \centering
        \caption{Spectrum of chiral particles of Model~\ref{tab: model_001_4}}
    \begin{tabular}{|c||c||c|c|c|c||c|c|}
    \hline
        Model~\ref{tab: model_001_4}  & $\SpecialUnitaryGroup{4}_C \times \SpecialUnitaryGroup{4}_L \times \SpecialUnitaryGroup{2}_R \times \mathrm{USp}(2)$ & $Q_{4C}$& $Q_{4L}$ & $Q_{2R}$ & $Q_{em}$ & $B-L$ & Field \\
        \hline \hline
        $ab$ & $4\times (4, \overline{4}, 1, 1)$ & $1$ & $-1$ & $0$ & $-\frac{1}{3}, \frac{2}{3}, -1, 0$ & $\frac{1}{3}, -1$& $Q_L, L_L$\\
        $ac$ & $4 \times (\overline{4} ,1 ,2, 1)$ & $-1$ &$0$ & $1$ & $-\frac{1}{3}, \frac{2}{3}, -1, 0$ & $\frac{1}{3}, -1$ & $Q_R, L_R$ \\
        $bc'$ & $4 \times (1 ,4 ,2, 1)$ & $0$ &$1$ & $1$ & $0, \pm 1$ & $0$ & $H'$ \\
        $b1$ & $4 \times (1 ,2 ,1, \overline{2})$ & $0$ &$1$ & $0$ & $\pm \frac{1}{2}$ & $0$ &  \\
        $b_{\Ysymm}$ & $2 \times (1, 10, 1, 1)$ & $0$ & $2$& 0& $\frac{1}{3}, -\frac{1}{3}, -1$ & $\frac{2}{3}, -2$ & \\
        $b_{\overline{\Yasymm}}$ & $2\times(1, \overline{6}, 1, 1)$ & $0$ & $-2$ & $0$ &$-\frac{1}{3}, 1$ & $-\frac{2}{3}, 2$ & \\
         $c_{\overline{\Ysymm}}$ & $3\times (1, 1, \overline{3}, 1)$ & $0$ & $0$ & $2$ & $0, \pm 1$ & 0 & \\
         $c_{\Yasymm}$ & $3 \times (1, 1, 1, 1)$ & $0$ & $0$& $2$ & $0$ & $0$ & \\
          \hline\hline 
         $bc$ & $ 6 \times (1, 2,\overline{2}, 1)$ & $0$ &$1$ & $-1$ & $0, \pm 1$ & $0$ & $H^i_u, H^i_d$\\
              & $6 \times (1, \overline{2}, 2, 1)$ & $0$ &$-1$ & $1$ &  &  & \\
        \hline
    \end{tabular}
    \label{tab: spectrum_model_001_4}
\end{table}

For the models with one tilted torus, the models represented by Model~\ref{tab: model_001_1} are the only class with gauge coupling unification and carry two confining $\mathrm{USp}$ groups. Thus gaugino condensation can trigger supersymmetry breaking and moduli stabilization \cite{cveticDynamicalSupersymmetryBreaking2003}.

\section{Discussions and Conclusions}\label{sec: conclusion}

Utilizing the deterministic algorithm,  we obtain various classes of  four-family supersymmetric models from intersecting $D6\mhyphen$branes on $\Quotient{\Torus{6}}{\Integer_2 \times\Integer_2}$ orientifold, with and without tilted torus. 
In total, there are $274$ physical independent four-family supersymmetric models without tilted torus, and $6$ physical independent four-family supersymmetric models with the third torus to be the tilted one, without loss of generality.  

For models without tilted torus, Model~\ref{tab: model_000_1} represents the model with gauge coupling unification at the string scale. Models~\ref{tab: model_000_2} and \ref{tab: model_000_3} are the rare models without any $\mathrm{USp}$ group in the hidden sectors, with tadpole cancellation conditions satisfied. Models~\ref{tab: model_000_9}-\ref{tab: model_000_13} are with at least two confining $\mathrm{USp}$ groups. Thus gaugino condensation can be triggered to break the supersymmetry and stabilize the moduli.
Moreover, there are Models~\ref{tab: model_000_14}-\ref{tab: model_000_22} with wrapping numbers absolutely larger than $5$ which was not reached for three family supersymmetric Pati-Salam models as discussed in~\cite{heCompleteSearchSupersymmetric2021}. 

For models with one tilted torus, Models~\ref{tab: model_001_2} and \ref{tab: model_001_3} satisfy the tadpole cancellation conditions without any filler branes. Models such as Models~\ref{tab: model_001_4} and \ref{tab: model_001_5} are related with $b$ and $c$ stacks of branes swapping. Model~\ref{tab: model_001_1} represents the models with exact gauge coupling unification at the string scale, with two confining $\mathrm{USp}$ groups allowing gaugino condensation as well. This class of models would be ideal for further phenomenology model buildings~(such as in~\cite{He:2021kbj}) as gaugino condensation can be triggered to break the supersymmetry and stabilize the moduli, while gauge coupling unified at string scale.

\acknowledgments
	TL and CZ are supported by the National Key Research and Development Program of China Grant No. 2020YFC2201504, 
by the Projects No. 11875062, No. 11947302, and No. 12047503 supported 
by the National Natural Science Foundation of China, as well as by the Key Research Program 
of the Chinese Academy of Sciences, Grant No. XDPB15. 
RS is supported by KIAS Individual Grant PG080701. RS would like to thank Weikun He for useful discussions. CZ would like to thank Lina Wu for helpful discussions.

\bibliography{Pati-Salam-Ref}
\newpage
\appendix
\section{Four-Family Standard Models from Intersecting D6-Branes without Tilted Tori}

In the appendix, we list all representative four-family models obtained from our random scanning method. In the first columns for each table, $a, b, c$ represent three stacks of $D6\mhyphen$branes, respectively. Also in the first columns, $1, 2, 3, 4$ is a short-handed notation for the filler branes along the $\Omega R$, $\Omega R \omega$, $\Omega R \theta \omega$ and $\Omega R \theta$ $O6\mhyphen$planes, respectively. The second columns for each table list the numbers of $D6\mhyphen$branes in every stack, respectively. In the third columns of each table, we record wrapping numbers of each $D6\mhyphen$brane configuration, and designate the third $\Torus{2}$ to be tilted.

The rest columns of each table record intersection numbers between stacks. For instance, in the $b$ column of Table~\ref{tab: model_001_2}, from top to bottom, the numbers represent intersection numbers $I_{ab}, I_{bb}, I_{cb}$, \textit{etc.}.  As usual, $b'$ and $c'$ are the orientifold $\Omega R$ image of $b$ and $c$ stacks of $D6\mhyphen$branes. We also list the relation between $x_A, x_B, x_C, x_D$, which are determined by the supersymmetry condition Equation~\eqref{eq: SUSY_condition}, as well as the relation between the moduli parameter $\chi_1, \chi_2, \chi_3$. The one loop $\beta$ functions $\beta^g_i$ are also listed. To have a clearer sight of gauge couplings, we list them up in the caption of each table, which makes it easier to check whether they are unified.

\begin{table}[h]\scriptsize
	\caption{D6-brane configurations and intersection numbers of Model \ref{tab: model_000_1}, and its gauge coupling relation is $g^2_a=g^2_b=g^2_c=(\frac{5}{3}g^2_Y)=\frac{8}{3} \sqrt[4]{2} \pi  e^{\phi ^4}$.}
	\label{tab: model_000_1}
	\begin{center}
		\begin{tabular}{|c||c|c||c|c|c|c|c|c|c|c|}
			\hline\rm{model} \ref{tab: model_000_1} & \multicolumn{10}{c|}{$U(4)\times U(2)_L\times U(2)_R\times USp(2)\times USp(4) $}\\
			\hline \hline			\rm{stack} & $N$ & $(n^1,l^1)\times(n^2,l^2)\times(n^3,l^3)$ & $n_{\Ysymm}$& $n_{\Yasymm}$ & $b$ & $b'$ & $c$ & $c'$ & 1 & 4\\
			\hline
			$a$ & 8 & $(-1,0)\times (-1,1)\times (1,2)$ & -1 & 1  & 4 & 0 & -4 & 0 & 0 & 2\\
			$b$ & 4 & $(0,1)\times (-1,2)\times (-1,2)$ & 3 & -3  & - & - & 0 & -8 & -4 & 0\\
			$c$ & 4 & $(1,1)\times (-1,0)\times (-1,2)$ & 1 & -1  & - & - & - & - & 0 & -2\\
			\hline
			1 & 2 & $(1, 0)\times (1, 0)\times (1, 0)$& \multicolumn{8}{c|}{$x_A = \frac{1}{4}x_B = \frac{1}{4}x_C = \frac{1}{2}x_D$}\\
			4 & 4 & $(0, 1)\times (0, 1)\times (1, 0)$& \multicolumn{8}{c|}{$\beta^g_1=-2$, $\beta^g_4=0$}\\
			& & & \multicolumn{8}{c|}{$\chi_1=\frac{1}{\sqrt{2}}$, $\chi_2=\frac{1}{\sqrt{2}}$, $\chi_3=\frac{1}{\sqrt{2}}$}\\
			\hline
		\end{tabular}
	\end{center}
\end{table}

\begin{table}[h]\scriptsize
	\caption{D6-brane configurations and intersection numbers of Model \ref{tab: model_000_2}, and its gauge coupling relation is $g^2_a=\frac{7}{9}g^2_b=\frac{7}{3}g^2_c=\frac{35}{23}(\frac{5}{3}g^2_Y)=\frac{4\ 5^{3/4} \pi  e^{\phi ^4}}{3 \sqrt{3}}$.}
	\label{tab: model_000_2}
	\begin{center}
		\begin{tabular}{|c||c|c||c|c|c|c|c|c|}
			\hline\rm{model} \ref{tab: model_000_2} & \multicolumn{8}{c|}{$U(4)\times U(2)_L\times U(2)_R  $}\\
			\hline \hline			\rm{stack} & $N$ & $(n^1,l^1)\times(n^2,l^2)\times(n^3,l^3)$ & $n_{\Ysymm}$& $n_{\Yasymm}$ & $b$ & $b'$ & $c$ & $c'$\\
			\hline
			$a$ & 8 & $(-1,0)\times (-1,1)\times (1,1)$ & 0 & 0  & 8 & -4 & -4 & 0\\
			$b$ & 4 & $(-1,2)\times (0,1)\times (-1,3)$ & 5 & -5  & - & - & -16 & 0\\
			$c$ & 4 & $(1,2)\times (-2,1)\times (-1,1)$ & -7 & -9  & - & - & - & -\\
			\hline
			& & & \multicolumn{6}{c|}{$x_A = \frac{1}{5}x_B = \frac{1}{6}x_C = \frac{1}{6}x_D$}\\
			& & & \multicolumn{6}{c|}{$\chi_1=\frac{\sqrt{5}}{6}$, $\chi_2=\frac{1}{\sqrt{5}}$, $\chi_3=\frac{2}{\sqrt{5}}$}\\
			\hline
		\end{tabular}
	\end{center}
\end{table}

\begin{table}\scriptsize
	\caption{D6-brane configurations and intersection numbers of Model \ref{tab: model_000_3}, and its gauge coupling relation is $g^2_a=\frac{10}{9}g^2_b=\frac{10}{3}g^2_c=\frac{50}{29}(\frac{5}{3}g^2_Y)=\frac{2}{3} \sqrt{\frac{2}{3}} 11^{3/4} \pi  e^{\phi ^4}$.}
	\label{tab: model_000_3}
	\begin{center}
		\begin{tabular}{|c||c|c||c|c|c|c|c|c|}
			\hline\rm{model} \ref{tab: model_000_3} & \multicolumn{8}{c|}{$U(4)\times U(4)_L\times U(4)_R  $}\\
			\hline \hline			\rm{stack} & $N$ & $(n^1,l^1)\times(n^2,l^2)\times(n^3,l^3)$ & $n_{\Ysymm}$& $n_{\Yasymm}$ & $b$ & $b'$ & $c$ & $c'$\\
			\hline
			$a$ & 8 & $(-1,0)\times (1,1)\times (-1,1)$ & 0 & 0  & 8 & -4 & -4 & 0\\
			$b$ & 8 & $(-1,1)\times (-1,3)\times (0,1)$ & 4 & -4  & - & - & 0 & -8\\
			$c$ & 8 & $(-1,1)\times (1,2)\times (-1,-1)$ & 8 & 24  & - & - & - & -\\
			\hline
			& & & \multicolumn{6}{c|}{$x_A = \frac{1}{11}x_B = \frac{1}{3}x_C = \frac{1}{3}x_D$}\\
			& & & \multicolumn{6}{c|}{$\chi_1=\frac{\sqrt{11}}{3}$, $\chi_2=\frac{1}{\sqrt{11}}$, $\chi_3=\frac{2}{\sqrt{11}}$}\\
			\hline
		\end{tabular}
	\end{center}
\end{table}

\begin{table}\scriptsize
	\caption{D6-brane configurations and intersection numbers of Model \ref{tab: model_000_4}, and its gauge coupling relation is $g^2_a=\frac{5}{6}g^2_b=\frac{5}{2}g^2_c=\frac{25}{16}(\frac{5}{3}g^2_Y)=\frac{4}{9} 14^{3/4} \pi  e^{\phi ^4}$.}
	\label{tab: model_000_4}
	\begin{center}
		\begin{tabular}{|c||c|c||c|c|c|c|c|c|c|}
			\hline\rm{model} \ref{tab: model_000_4} & \multicolumn{9}{c|}{$U(4)\times U(2)_L\times U(2)_R\times USp(2) $}\\
			\hline \hline			\rm{stack} & $N$ & $(n^1,l^1)\times(n^2,l^2)\times(n^3,l^3)$ & $n_{\Ysymm}$& $n_{\Yasymm}$ & $b$ & $b'$ & $c$ & $c'$ & 2\\
			\hline
			$a$ & 8 & $(-1,0)\times (-1,1)\times (1,1)$ & 0 & 0  & 6 & -2 & -4 & 0 & 0\\
			$b$ & 4 & $(-1,2)\times (0,1)\times (-1,2)$ & 3 & -3  & - & - & -8 & 0 & 0\\
			$c$ & 4 & $(1,2)\times (-2,1)\times (-1,1)$ & -7 & -9  & - & - & - & - & 4\\
			\hline
			2 & 2 & $(1, 0)\times (0, 1)\times (0, 1)$& \multicolumn{7}{c|}{$x_A = \frac{2}{7}x_B = \frac{1}{4}x_C = \frac{1}{4}x_D$}\\
			& & & \multicolumn{7}{c|}{$\beta^g_2=-2$}\\
			& & & \multicolumn{7}{c|}{$\chi_1=\frac{\sqrt{\frac{7}{2}}}{4}$, $\chi_2=\sqrt{\frac{2}{7}}$, $\chi_3=2 \sqrt{\frac{2}{7}}$}\\
			\hline
		\end{tabular}
	\end{center}
\end{table}

\begin{table}\scriptsize
	\caption{D6-brane configurations and intersection numbers of Model \ref{tab: model_000_5}, and its gauge coupling relation is $g^2_a=g^2_b=3g^2_c=\frac{5}{3}(\frac{5}{3}g^2_Y)=\frac{8}{3} 2^{3/4} \pi  e^{\phi ^4}$.}
	\label{tab: model_000_5}
	\begin{center}
		\begin{tabular}{|c||c|c||c|c|c|c|c|c|c|}
			\hline\rm{model} \ref{tab: model_000_5} & \multicolumn{9}{c|}{$U(4)\times U(4)_L\times U(4)_R\times USp(4) $}\\
			\hline \hline			\rm{stack} & $N$ & $(n^1,l^1)\times(n^2,l^2)\times(n^3,l^3)$ & $n_{\Ysymm}$& $n_{\Yasymm}$ & $b$ & $b'$ & $c$ & $c'$ & 3\\
			\hline
			$a$ & 8 & $(-1,0)\times (-1,2)\times (1,1)$ & 1 & -1  & 4 & 0 & -4 & 0 & -2\\
			$b$ & 8 & $(-1,1)\times (0,1)\times (-1,1)$ & 0 & 0  & - & - & 0 & 0 & 2\\
			$c$ & 8 & $(1,1)\times (-1,1)\times (-1,1)$ & -4 & -12  & - & - & - & - & -2\\
			\hline
			3 & 4 & $(0, 1)\times (1, 0)\times (0, 1)$& \multicolumn{7}{c|}{$x_A = \frac{1}{4}x_B = x_C = \frac{1}{2}x_D$}\\
			& & & \multicolumn{7}{c|}{$\beta^g_3=2$}\\
			& & & \multicolumn{7}{c|}{$\chi_1=\sqrt{2}$, $\chi_2=\frac{1}{2 \sqrt{2}}$, $\chi_3=\sqrt{2}$}\\
			\hline
		\end{tabular}
	\end{center}
\end{table}

\begin{table}\scriptsize
	\caption{D6-brane configurations and intersection numbers of Model \ref{tab: model_000_6}, and its gauge coupling relation is $g^2_a=2g^2_b=g^2_c=(\frac{5}{3}g^2_Y)=\frac{8}{3} 2^{3/4} \pi  e^{\phi ^4}$.}
	\label{tab: model_000_6}
	\begin{center}
		\begin{tabular}{|c||c|c||c|c|c|c|c|c|c|c|}
			\hline\rm{model} \ref{tab: model_000_6} & \multicolumn{10}{c|}{$U(4)\times U(4)_L\times U(2)_R\times USp(2)\times USp(4) $}\\
			\hline \hline			\rm{stack} & $N$ & $(n^1,l^1)\times(n^2,l^2)\times(n^3,l^3)$ & $n_{\Ysymm}$& $n_{\Yasymm}$ & $b$ & $b'$ & $c$ & $c'$ & 1 & 4\\
			\hline
			$a$ & 8 & $(-1,0)\times (-1,1)\times (1,2)$ & -1 & 1  & 0 & 4 & -4 & 0 & 0 & 2\\
			$b$ & 8 & $(0,1)\times (-1,1)\times (-1,1)$ & 0 & 0  & - & - & 2 & -6 & -2 & 0\\
			$c$ & 4 & $(1,1)\times (-1,0)\times (-1,2)$ & 1 & -1  & - & - & - & - & 0 & -2\\
			\hline
			1 & 2 & $(1, 0)\times (1, 0)\times (1, 0)$& \multicolumn{8}{c|}{$x_A = x_B = x_C = 2x_D$}\\
			4 & 4 & $(0, 1)\times (0, 1)\times (1, 0)$& \multicolumn{8}{c|}{$\beta^g_1=-4$, $\beta^g_4=0$}\\
			& & & \multicolumn{8}{c|}{$\chi_1=\sqrt{2}$, $\chi_2=\sqrt{2}$, $\chi_3=\sqrt{2}$}\\
			\hline
		\end{tabular}
	\end{center}
\end{table}

\begin{table}\scriptsize
	\caption{D6-brane configurations and intersection numbers of Model \ref{tab: model_000_7}, and its gauge coupling relation is $g^2_a=\frac{21}{22}g^2_b=\frac{7}{2}g^2_c=\frac{7}{4}(\frac{5}{3}g^2_Y)=\frac{4}{11} \sqrt{3} 10^{3/4} \pi  e^{\phi ^4}$.}
	\label{tab: model_000_7}
	\begin{center}
		\begin{tabular}{|c||c|c||c|c|c|c|c|c|c|c|}
			\hline\rm{model} \ref{tab: model_000_7} & \multicolumn{10}{c|}{$U(4)\times U(2)_L\times U(2)_R\times USp(4)^2 $}\\
			\hline \hline			\rm{stack} & $N$ & $(n^1,l^1)\times(n^2,l^2)\times(n^3,l^3)$ & $n_{\Ysymm}$& $n_{\Yasymm}$ & $b$ & $b'$ & $c$ & $c'$ & 2 & 4\\
			\hline
			$a$ & 8 & $(-1,0)\times (1,1)\times (-1,2)$ & 1 & -1  & 6 & -2 & -4 & 0 & 0 & -2\\
			$b$ & 4 & $(-3,2)\times (-1,2)\times (0,1)$ & 1 & -1  & - & - & -4 & 0 & 0 & 3\\
			$c$ & 4 & $(-2,1)\times (1,2)\times (-1,-2)$ & 5 & 27  & - & - & - & - & -1 & 4\\
			\hline
			2 & 4 & $(1, 0)\times (0, 1)\times (0, 1)$& \multicolumn{8}{c|}{$x_A = \frac{1}{20}x_B = \frac{3}{8}x_C = \frac{3}{4}x_D$}\\
			4 & 4 & $(0, 1)\times (0, 1)\times (1, 0)$& \multicolumn{8}{c|}{$\beta^g_2=-5$, $\beta^g_4=5$}\\
			& & & \multicolumn{8}{c|}{$\chi_1=\frac{3 \sqrt{\frac{5}{2}}}{2}$, $\chi_2=\frac{1}{\sqrt{10}}$, $\chi_3=\frac{1}{\sqrt{10}}$}\\
			\hline
		\end{tabular}
	\end{center}
\end{table}

\begin{table}\scriptsize
	\caption{D6-brane configurations and intersection numbers of Model \ref{tab: model_000_8}, and its gauge coupling relation is $g^2_a=g^2_b=4g^2_c=\frac{20}{11}(\frac{5}{3}g^2_Y)=\frac{32}{9} \sqrt[4]{2} \pi  e^{\phi ^4}$.}
	\label{tab: model_000_8}
	\begin{center}
		\begin{tabular}{|c||c|c||c|c|c|c|c|c|c|c|}
			\hline\rm{model} \ref{tab: model_000_8} & \multicolumn{10}{c|}{$U(4)\times U(4)_L\times U(2)_R\times USp(4)\times USp(12) $}\\
			\hline \hline			\rm{stack} & $N$ & $(n^1,l^1)\times(n^2,l^2)\times(n^3,l^3)$ & $n_{\Ysymm}$& $n_{\Yasymm}$ & $b$ & $b'$ & $c$ & $c'$ & 2 & 4\\
			\hline
			$a$ & 8 & $(-1,0)\times (1,1)\times (-1,2)$ & 1 & -1  & 4 & 0 & -4 & 0 & 0 & -2\\
			$b$ & 8 & $(-1,1)\times (-1,1)\times (0,1)$ & 0 & 0  & - & - & -6 & 6 & 0 & 2\\
			$c$ & 4 & $(-2,1)\times (1,2)\times (-1,-2)$ & 5 & 27  & - & - & - & - & -1 & 4\\
			\hline
			2 & 12 & $(1, 0)\times (0, 1)\times (0, 1)$& \multicolumn{8}{c|}{$x_A = \frac{1}{16}x_B = \frac{1}{2}x_C = x_D$}\\
			4 & 4 & $(0, 1)\times (0, 1)\times (1, 0)$& \multicolumn{8}{c|}{$\beta^g_2=-5$, $\beta^g_4=4$}\\
			& & & \multicolumn{8}{c|}{$\chi_1=2 \sqrt{2}$, $\chi_2=\frac{1}{2 \sqrt{2}}$, $\chi_3=\frac{1}{2 \sqrt{2}}$}\\
			\hline
		\end{tabular}
	\end{center}
\end{table}

\begin{table}\scriptsize
	\caption{D6-brane configurations and intersection numbers of Model \ref{tab: model_000_9}, and its gauge coupling relation is $g^2_a=g^2_b=2g^2_c=\frac{10}{7}(\frac{5}{3}g^2_Y)=2 \sqrt{2} \sqrt[4]{3} \pi  e^{\phi ^4}$.}
	\label{tab: model_000_9}
	\begin{center}
		\begin{tabular}{|c||c|c||c|c|c|c|c|c|c|c|c|}
			\hline\rm{model} \ref{tab: model_000_9} & \multicolumn{11}{c|}{$U(4)\times U(2)_L\times U(2)_R\times USp(2)^2\times USp(4) $}\\
			\hline \hline			\rm{stack} & $N$ & $(n^1,l^1)\times(n^2,l^2)\times(n^3,l^3)$ & $n_{\Ysymm}$& $n_{\Yasymm}$ & $b$ & $b'$ & $c$ & $c'$ & 2 & 3 & 4\\
			\hline
			$a$ & 8 & $(-1,0)\times (-1,1)\times (1,1)$ & 0 & 0  & 4 & 0 & -4 & 0 & 0 & -1 & 1\\
			$b$ & 4 & $(0,1)\times (-1,3)\times (-1,1)$ & 2 & -2  & - & - & 0 & -12 & 1 & 0 & 0\\
			$c$ & 4 & $(1,1)\times (-1,0)\times (-2,2)$ & 0 & 0  & - & - & - & - & 2 & 0 & -2\\
			\hline
			2 & 4 & $(1, 0)\times (0, 1)\times (0, 1)$& \multicolumn{9}{c|}{$x_A = \frac{1}{3}x_B = \frac{1}{3}x_C = \frac{1}{3}x_D$}\\
			3 & 2 & $(0, 1)\times (1, 0)\times (0, 1)$& \multicolumn{9}{c|}{$\beta^g_2=-3$, $\beta^g_3=-4$, $\beta^g_4=-2$}\\
			4 & 2 & $(0, 1)\times (0, 1)\times (1, 0)$& \multicolumn{9}{c|}{$\chi_1=\frac{1}{\sqrt{3}}$, $\chi_2=\frac{1}{\sqrt{3}}$, $\chi_3=\frac{2}{\sqrt{3}}$}\\
			\hline
		\end{tabular}
	\end{center}
\end{table}

\begin{table}\scriptsize
	\caption{D6-brane configurations and intersection numbers of Model \ref{tab: model_000_10}, and its gauge coupling relation is $g^2_a=g^2_b=\frac{5}{2}g^2_c=\frac{25}{16}(\frac{5}{3}g^2_Y)=\frac{2}{3} 11^{3/4} \pi  e^{\phi ^4}$.}
	\label{tab: model_000_10}
	\begin{center}
		\begin{tabular}{|c||c|c||c|c|c|c|c|c|c|c|c|c|}
			\hline\rm{model} \ref{tab: model_000_10} & \multicolumn{12}{c|}{$U(4)\times U(2)_L\times U(2)_R\times USp(2)^3\times USp(8) $}\\
			\hline \hline			\rm{stack} & $N$ & $(n^1,l^1)\times(n^2,l^2)\times(n^3,l^3)$ & $n_{\Ysymm}$& $n_{\Yasymm}$ & $b$ & $b'$ & $c$ & $c'$ & 1 & 2 & 3 & 4\\
			\hline
			$a$ & 8 & $(-1,0)\times (1,1)\times (-1,1)$ & 0 & 0  & 4 & 0 & -4 & 0 & 0 & 0 & 1 & -1\\
			$b$ & 4 & $(-1,2)\times (-1,1)\times (0,1)$ & 1 & -1  & - & - & -4 & 6 & -2 & 0 & 0 & 1\\
			$c$ & 4 & $(-1,1)\times (1,3)\times (-1,-1)$ & 0 & 12  & - & - & - & - & 3 & -1 & 3 & 1\\
			\hline
			1 & 2 & $(1, 0)\times (1, 0)\times (1, 0)$& \multicolumn{10}{c|}{$x_A = \frac{1}{11}x_B = \frac{1}{2}x_C = \frac{1}{2}x_D$}\\
			2 & 8 & $(1, 0)\times (0, 1)\times (0, 1)$& \multicolumn{10}{c|}{$\beta^g_1=-1$, $\beta^g_2=-5$, $\beta^g_3=-1$, $\beta^g_4=-2$}\\
			3 & 2 & $(0, 1)\times (1, 0)\times (0, 1)$& \multicolumn{10}{c|}{$\chi_1=\frac{\sqrt{11}}{2}$, $\chi_2=\frac{1}{\sqrt{11}}$, $\chi_3=\frac{2}{\sqrt{11}}$}\\
			4 & 2 & $(0, 1)\times (0, 1)\times (1, 0)$& \multicolumn{10}{c|}{}\\
			\hline
		\end{tabular}
	\end{center}
\end{table}

\begin{table}\scriptsize
	\caption{D6-brane configurations and intersection numbers of Model \ref{tab: model_000_11}, and its gauge coupling relation is $g^2_a=\frac{3}{2}g^2_b=g^2_c=(\frac{5}{3}g^2_Y)=2\ 3^{3/4} \pi  e^{\phi ^4}$.}
	\label{tab: model_000_11}
	\begin{center}
		\begin{tabular}{|c||c|c||c|c|c|c|c|c|c|c|c|c|}
			\hline\rm{model} \ref{tab: model_000_11} & \multicolumn{12}{c|}{$U(4)\times U(2)_L\times U(2)_R\times USp(2)^3\times USp(4) $}\\
			\hline \hline			\rm{stack} & $N$ & $(n^1,l^1)\times(n^2,l^2)\times(n^3,l^3)$ & $n_{\Ysymm}$& $n_{\Yasymm}$ & $b$ & $b'$ & $c$ & $c'$ & 1 & 2 & 3 & 4\\
			\hline
			$a$ & 8 & $(-1,0)\times (-1,1)\times (1,1)$ & 0 & 0  & 0 & 4 & -4 & 0 & 0 & 0 & -1 & 1\\
			$b$ & 4 & $(0,1)\times (-1,1)\times (-3,1)$ & -2 & 2  & - & - & 2 & -4 & -1 & 3 & 0 & 0\\
			$c$ & 4 & $(1,2)\times (-1,0)\times (-1,1)$ & -1 & 1  & - & - & - & - & 0 & 2 & 0 & -1\\
			\hline
			1 & 2 & $(1, 0)\times (1, 0)\times (1, 0)$& \multicolumn{10}{c|}{$x_A = 3x_B = \frac{3}{2}x_C = \frac{3}{2}x_D$}\\
			2 & 4 & $(1, 0)\times (0, 1)\times (0, 1)$& \multicolumn{10}{c|}{$\beta^g_1=-5$, $\beta^g_2=-1$, $\beta^g_3=-4$, $\beta^g_4=-3$}\\
			3 & 2 & $(0, 1)\times (1, 0)\times (0, 1)$& \multicolumn{10}{c|}{$\chi_1=\frac{\sqrt{3}}{2}$, $\chi_2=\sqrt{3}$, $\chi_3=2 \sqrt{3}$}\\
			4 & 2 & $(0, 1)\times (0, 1)\times (1, 0)$& \multicolumn{10}{c|}{}\\
			\hline
		\end{tabular}
	\end{center}
\end{table}

\begin{table}\scriptsize
	\caption{D6-brane configurations and intersection numbers of Model \ref{tab: model_000_12}, and its gauge coupling relation is $g^2_a=g^2_b=\frac{3}{2}g^2_c=\frac{5}{4}(\frac{5}{3}g^2_Y)=2\ 3^{3/4} \pi  e^{\phi ^4}$.}
	\label{tab: model_000_12}
	\begin{center}
		\begin{tabular}{|c||c|c||c|c|c|c|c|c|c|c|c|c|}
			\hline\rm{model} \ref{tab: model_000_12} & \multicolumn{12}{c|}{$U(4)\times U(2)_L\times U(2)_R\times USp(2)^3\times USp(4) $}\\
			\hline \hline			\rm{stack} & $N$ & $(n^1,l^1)\times(n^2,l^2)\times(n^3,l^3)$ & $n_{\Ysymm}$& $n_{\Yasymm}$ & $b$ & $b'$ & $c$ & $c'$ & 1 & 2 & 3 & 4\\
			\hline
			$a$ & 8 & $(1,0)\times (1,1)\times (1,-1)$ & 0 & 0  & 4 & 0 & -4 & 0 & 0 & 0 & 1 & -1\\
			$b$ & 4 & $(-1,2)\times (-1,1)\times (0,1)$ & 1 & -1  & - & - & 4 & 2 & -2 & 0 & 0 & 1\\
			$c$ & 4 & $(0,1)\times (1,3)\times (-1,-1)$ & -2 & 2  & - & - & - & - & 3 & -1 & 0 & 0\\
			\hline
			1 & 4 & $(1, 0)\times (1, 0)\times (1, 0)$& \multicolumn{10}{c|}{$x_A = \frac{1}{3}x_B = \frac{1}{2}x_C = \frac{1}{2}x_D$}\\
			2 & 2 & $(1, 0)\times (0, 1)\times (0, 1)$& \multicolumn{10}{c|}{$\beta^g_1=-1$, $\beta^g_2=-5$, $\beta^g_3=-4$, $\beta^g_4=-3$}\\
			3 & 2 & $(0, 1)\times (1, 0)\times (0, 1)$& \multicolumn{10}{c|}{$\chi_1=\frac{\sqrt{3}}{2}$, $\chi_2=\frac{1}{\sqrt{3}}$, $\chi_3=\frac{2}{\sqrt{3}}$}\\
			4 & 2 & $(0, 1)\times (0, 1)\times (1, 0)$& \multicolumn{10}{c|}{}\\
			\hline
		\end{tabular}
	\end{center}
\end{table}

\begin{table}\scriptsize
	\caption{D6-brane configurations and intersection numbers of Model \ref{tab: model_000_13}, and its gauge coupling relation is $g^2_a=\frac{11}{10}g^2_b=\frac{11}{5}g^2_c=\frac{55}{37}(\frac{5}{3}g^2_Y)=\frac{8\ 13^{3/4} \pi  e^{\phi ^4}}{7 \sqrt{5}}$.}
	\label{tab: model_000_13}
	\begin{center}
		\begin{tabular}{|c||c|c||c|c|c|c|c|c|c|c|c|}
			\hline\rm{model} \ref{tab: model_000_13} & \multicolumn{11}{c|}{$U(4)\times U(2)_L\times U(2)_R\times USp(2)^3 $}\\
			\hline \hline			\rm{stack} & $N$ & $(n^1,l^1)\times(n^2,l^2)\times(n^3,l^3)$ & $n_{\Ysymm}$& $n_{\Yasymm}$ & $b$ & $b'$ & $c$ & $c'$ & 2 & 3 & 4\\
			\hline
			$a$ & 8 & $(1,-1)\times (1,0)\times (1,1)$ & 0 & 0  & -3 & 7 & -4 & 0 & -1 & 0 & 1\\
			$b$ & 4 & $(-5,2)\times (1,1)\times (-1,0)$ & -3 & 3  & - & - & -2 & 0 & -2 & 5 & 0\\
			$c$ & 4 & $(-3,1)\times (-1,1)\times (-1,1)$ & 0 & 12  & - & - & - & - & 1 & 3 & 3\\
			\hline
			2 & 2 & $(1, 0)\times (0, 1)\times (0, 1)$& \multicolumn{9}{c|}{$x_A = \frac{26}{5}x_B = 13x_C = \frac{26}{5}x_D$}\\
			3 & 2 & $(0, 1)\times (1, 0)\times (0, 1)$& \multicolumn{9}{c|}{$\beta^g_2=-1$, $\beta^g_3=2$, $\beta^g_4=-1$}\\
			4 & 2 & $(0, 1)\times (0, 1)\times (1, 0)$& \multicolumn{9}{c|}{$\chi_1=\sqrt{13}$, $\chi_2=\frac{2 \sqrt{13}}{5}$, $\chi_3=2 \sqrt{13}$}\\
			\hline
		\end{tabular}
	\end{center}
\end{table}

\begin{table}\scriptsize
	\caption{D6-brane configurations and intersection numbers of Model \ref{tab: model_000_14}, and its gauge coupling relation is $g^2_a=\frac{6}{7}g^2_b=\frac{2}{7}g^2_c=\frac{2}{5}(\frac{5}{3}g^2_Y)=\frac{6}{7} \sqrt{2} \sqrt[4]{3} \pi  e^{\phi ^4}$.}
	\label{tab: model_000_14}
	\begin{center}
		\begin{tabular}{|c||c|c||c|c|c|c|c|c|}
			\hline\rm{model} \ref{tab: model_000_14} & \multicolumn{8}{c|}{$U(4)\times U(2)_L\times U(2)_R  $}\\
			\hline \hline			\rm{stack} & $N$ & $(n^1,l^1)\times(n^2,l^2)\times(n^3,l^3)$ & $n_{\Ysymm}$& $n_{\Yasymm}$ & $b$ & $b'$ & $c$ & $c'$\\
			\hline
			$a$ & 8 & $(-2,-1)\times (1,1)\times (1,1)$ & 0 & -8  & 12 & -8 & -4 & 0\\
			$b$ & 4 & $(-1,1)\times (6,2)\times (-1,0)$ & 4 & -4  & - & - & 8 & 0\\
			$c$ & 4 & $(1,1)\times (-1,0)\times (-2,2)$ & 0 & 0  & - & - & - & -\\
			\hline
			& & & \multicolumn{6}{c|}{$x_A = 9x_B = 3x_C = 9x_D$}\\
			& & & \multicolumn{6}{c|}{$\chi_1=\sqrt{3}$, $\chi_2=3 \sqrt{3}$, $\chi_3=2 \sqrt{3}$}\\
			\hline
		\end{tabular}
	\end{center}
\end{table}

\begin{table}\scriptsize
	\caption{D6-brane configurations and intersection numbers of Model \ref{tab: model_000_15}, and its gauge coupling relation is $g^2_a=\frac{47}{112}g^2_b=\frac{18}{7}g^2_c=\frac{30}{19}(\frac{5}{3}g^2_Y)=\frac{1}{3} \sqrt{\frac{2}{7}} 23^{3/4} \pi  e^{\phi ^4}$.}
	\label{tab: model_000_15}
	\begin{center}
		\begin{tabular}{|c||c|c||c|c|c|c|c|c|c|c|}
			\hline\rm{model} \ref{tab: model_000_15} & \multicolumn{10}{c|}{$U(4)\times U(2)_L\times U(2)_R\times USp(2)\times USp(4) $}\\
			\hline \hline			\rm{stack} & $N$ & $(n^1,l^1)\times(n^2,l^2)\times(n^3,l^3)$ & $n_{\Ysymm}$& $n_{\Yasymm}$ & $b$ & $b'$ & $c$ & $c'$ & 1 & 2\\
			\hline
			$a$ & 8 & $(1,-1)\times (1,0)\times (2,1)$ & 1 & -1  & -5 & 9 & -4 & 0 & 0 & -2\\
			$b$ & 4 & $(-7,2)\times (1,1)\times (-1,0)$ & -5 & 5  & - & - & 9 & 11 & 0 & -2\\
			$c$ & 4 & $(-2,1)\times (-2,1)\times (-2,1)$ & 5 & 27  & - & - & - & - & -1 & 4\\
			\hline
			1 & 2 & $(1, 0)\times (1, 0)\times (1, 0)$& \multicolumn{8}{c|}{$x_A = \frac{92}{7}x_B = 46x_C = \frac{46}{7}x_D$}\\
			2 & 4 & $(1, 0)\times (0, 1)\times (0, 1)$& \multicolumn{8}{c|}{$\beta^g_1=-5$, $\beta^g_2=4$}\\
			& & & \multicolumn{8}{c|}{$\chi_1=\sqrt{23}$, $\chi_2=\frac{2 \sqrt{23}}{7}$, $\chi_3=4 \sqrt{23}$}\\
			\hline
		\end{tabular}
	\end{center}
\end{table}

\begin{table}\scriptsize
	\caption{D6-brane configurations and intersection numbers of Model \ref{tab: model_000_16}, and its gauge coupling relation is $g^2_a=\frac{5}{24}g^2_b=\frac{19}{8}g^2_c=\frac{95}{62}(\frac{5}{3}g^2_Y)=\frac{2}{45} 86^{3/4} \pi  e^{\phi ^4}$.}
	\label{tab: model_000_16}
	\begin{center}
		\begin{tabular}{|c||c|c||c|c|c|c|c|c|c|c|}
			\hline\rm{model} \ref{tab: model_000_16} & \multicolumn{10}{c|}{$U(4)\times U(2)_L\times U(2)_R\times USp(2)^2 $}\\
			\hline \hline			\rm{stack} & $N$ & $(n^1,l^1)\times(n^2,l^2)\times(n^3,l^3)$ & $n_{\Ysymm}$& $n_{\Yasymm}$ & $b$ & $b'$ & $c$ & $c'$ & 2 & 4\\
			\hline
			$a$ & 8 & $(2,-1)\times (1,0)\times (2,1)$ & 0 & 0  & -6 & 10 & -4 & 0 & -2 & 2\\
			$b$ & 4 & $(-8,1)\times (1,1)\times (-1,0)$ & -7 & 7  & - & - & 15 & 11 & -1 & 0\\
			$c$ & 4 & $(-3,1)\times (-2,1)\times (-2,1)$ & 9 & 39  & - & - & - & - & 4 & 6\\
			\hline
			2 & 2 & $(1, 0)\times (0, 1)\times (0, 1)$& \multicolumn{8}{c|}{$x_A = \frac{43}{4}x_B = 86x_C = \frac{43}{4}x_D$}\\
			4 & 2 & $(0, 1)\times (0, 1)\times (1, 0)$& \multicolumn{8}{c|}{$\beta^g_2=3$, $\beta^g_4=4$}\\
			& & & \multicolumn{8}{c|}{$\chi_1=\sqrt{86}$, $\chi_2=\frac{\sqrt{\frac{43}{2}}}{4}$, $\chi_3=2 \sqrt{86}$}\\
			\hline
		\end{tabular}
	\end{center}
\end{table}

\begin{table}\scriptsize
	\caption{D6-brane configurations and intersection numbers of Model \ref{tab: model_000_17}, and its gauge coupling relation is $g^2_a=\frac{760}{31}g^2_b=9g^2_c=\frac{15}{7}(\frac{5}{3}g^2_Y)=\frac{48}{31} \sqrt{2} \sqrt[4]{3} 7^{3/4} \pi  e^{\phi ^4}$.}
	\label{tab: model_000_17}
	\begin{center}
		\begin{tabular}{|c||c|c||c|c|c|c|c|c|c|c|}
			\hline\rm{model} \ref{tab: model_000_17} & \multicolumn{10}{c|}{$U(4)\times U(2)_L\times U(2)_R\times USp(2)^2 $}\\
			\hline \hline			\rm{stack} & $N$ & $(n^1,l^1)\times(n^2,l^2)\times(n^3,l^3)$ & $n_{\Ysymm}$& $n_{\Yasymm}$ & $b$ & $b'$ & $c$ & $c'$ & 1 & 3\\
			\hline
			$a$ & 8 & $(-1,1)\times (-1,0)\times (1,1)$ & 0 & 0  & -5 & 9 & -4 & 0 & 0 & 0\\
			$b$ & 4 & $(-2,1)\times (-1,1)\times (-4,1)$ & 3 & 29  & - & - & 33 & -35 & -1 & 8\\
			$c$ & 4 & $(1,0)\times (-9,-2)\times (-1,1)$ & 7 & -7  & - & - & - & - & 0 & 2\\
			\hline
			1 & 2 & $(1, 0)\times (1, 0)\times (1, 0)$& \multicolumn{8}{c|}{$x_A = 42x_B = \frac{28}{3}x_C = 42x_D$}\\
			3 & 2 & $(0, 1)\times (1, 0)\times (0, 1)$& \multicolumn{8}{c|}{$\beta^g_1=-5$, $\beta^g_3=4$}\\
			& & & \multicolumn{8}{c|}{$\chi_1=2 \sqrt{\frac{7}{3}}$, $\chi_2=3 \sqrt{21}$, $\chi_3=4 \sqrt{\frac{7}{3}}$}\\
			\hline
		\end{tabular}
	\end{center}
\end{table}

\begin{table}\scriptsize
	\caption{D6-brane configurations and intersection numbers of Model \ref{tab: model_000_18}, and its gauge coupling relation is $g^2_a=\frac{1309}{47}g^2_b=5g^2_c=\frac{25}{13}(\frac{5}{3}g^2_Y)=\frac{8}{47} \sqrt[4]{2} 185^{3/4} \pi  e^{\phi ^4}$.}
	\label{tab: model_000_18}
	\begin{center}
		\begin{tabular}{|c||c|c||c|c|c|c|c|c|}
			\hline\rm{model} \ref{tab: model_000_18} & \multicolumn{8}{c|}{$U(4)\times U(2)_L\times U(2)_R  $}\\
			\hline \hline			\rm{stack} & $N$ & $(n^1,l^1)\times(n^2,l^2)\times(n^3,l^3)$ & $n_{\Ysymm}$& $n_{\Yasymm}$ & $b$ & $b'$ & $c$ & $c'$\\
			\hline
			$a$ & 8 & $(0,1)\times (-1,-2)\times (2,1)$ & 0 & 0  & 10 & -6 & -4 & 0\\
			$b$ & 4 & $(1,-2)\times (1,-3)\times (4,1)$ & -23 & -73  & - & - & 48 & -24\\
			$c$ & 4 & $(1,-10)\times (0,-1)\times (-2,1)$ & 8 & -8  & - & - & - & -\\
			\hline
			& & & \multicolumn{6}{c|}{$x_A = x_B = \frac{1}{5}x_C = \frac{1}{74}x_D$}\\
			& & & \multicolumn{6}{c|}{$\chi_1=\frac{1}{\sqrt{370}}$, $\chi_2=\sqrt{\frac{5}{74}}$, $\chi_3=2 \sqrt{\frac{74}{5}}$}\\
			\hline
		\end{tabular}
	\end{center}
\end{table}

\begin{table}\scriptsize
	\caption{D6-brane configurations and intersection numbers of Model \ref{tab: model_000_19}, and its gauge coupling relation is $g^2_a=\frac{1248}{41}g^2_b=11g^2_c=\frac{11}{5}(\frac{5}{3}g^2_Y)=\frac{64}{123} 77^{3/4} \pi  e^{\phi ^4}$.}
	\label{tab: model_000_19}
	\begin{center}
		\begin{tabular}{|c||c|c||c|c|c|c|c|c|c|}
			\hline\rm{model} \ref{tab: model_000_19} & \multicolumn{9}{c|}{$U(4)\times U(2)_L\times U(2)_R\times USp(14) $}\\
			\hline \hline			\rm{stack} & $N$ & $(n^1,l^1)\times(n^2,l^2)\times(n^3,l^3)$ & $n_{\Ysymm}$& $n_{\Yasymm}$ & $b$ & $b'$ & $c$ & $c'$ & 1\\
			\hline
			$a$ & 8 & $(-1,1)\times (-1,0)\times (1,1)$ & 0 & 0  & -5 & 9 & -4 & 0 & 0\\
			$b$ & 4 & $(-2,1)\times (-2,1)\times (-4,1)$ & 13 & 51  & - & - & 45 & -35 & -1\\
			$c$ & 4 & $(1,0)\times (-11,-2)\times (-1,1)$ & 9 & -9  & - & - & - & - & 0\\
			\hline
			1 & 14 & $(1, 0)\times (1, 0)\times (1, 0)$& \multicolumn{7}{c|}{$x_A = 56x_B = \frac{112}{11}x_C = 56x_D$}\\
			& & & \multicolumn{7}{c|}{$\beta^g_1=-5$}\\
			& & & \multicolumn{7}{c|}{$\chi_1=4 \sqrt{\frac{7}{11}}$, $\chi_2=2 \sqrt{77}$, $\chi_3=8 \sqrt{\frac{7}{11}}$}\\
			\hline
		\end{tabular}
	\end{center}
\end{table}

\begin{table}\scriptsize
	\caption{D6-brane configurations and intersection numbers of Model \ref{tab: model_000_20}, and its gauge coupling relation is $g^2_a=\frac{1680}{47}g^2_b=13g^2_c=\frac{65}{29}(\frac{5}{3}g^2_Y)=\frac{256}{141} \sqrt[4]{2} 13^{3/4} \pi  e^{\phi ^4}$.}
	\label{tab: model_000_20}
	\begin{center}
		\begin{tabular}{|c||c|c||c|c|c|c|c|c|c|}
			\hline\rm{model} \ref{tab: model_000_20} & \multicolumn{9}{c|}{$U(4)\times U(2)_L\times U(2)_R\times USp(10) $}\\
			\hline \hline			\rm{stack} & $N$ & $(n^1,l^1)\times(n^2,l^2)\times(n^3,l^3)$ & $n_{\Ysymm}$& $n_{\Yasymm}$ & $b$ & $b'$ & $c$ & $c'$ & 1\\
			\hline
			$a$ & 8 & $(-1,1)\times (-1,0)\times (1,1)$ & 0 & 0  & -5 & 9 & -4 & 0 & 0\\
			$b$ & 4 & $(-2,1)\times (-2,1)\times (-4,1)$ & 13 & 51  & - & - & 51 & -45 & -1\\
			$c$ & 4 & $(1,0)\times (-13,-2)\times (-1,1)$ & 11 & -11  & - & - & - & - & 0\\
			\hline
			1 & 10 & $(1, 0)\times (1, 0)\times (1, 0)$& \multicolumn{7}{c|}{$x_A = 64x_B = \frac{128}{13}x_C = 64x_D$}\\
			& & & \multicolumn{7}{c|}{$\beta^g_1=-5$}\\
			& & & \multicolumn{7}{c|}{$\chi_1=8 \sqrt{\frac{2}{13}}$, $\chi_2=4 \sqrt{26}$, $\chi_3=16 \sqrt{\frac{2}{13}}$}\\
			\hline
		\end{tabular}
	\end{center}
\end{table}

\begin{table}\scriptsize
	\caption{D6-brane configurations and intersection numbers of Model \ref{tab: model_000_21}, and its gauge coupling relation is $g^2_a=\frac{2176}{53}g^2_b=15g^2_c=\frac{25}{11}(\frac{5}{3}g^2_Y)=\frac{192}{53} \sqrt[4]{3} 5^{3/4} \pi  e^{\phi ^4}$.}
	\label{tab: model_000_21}
	\begin{center}
		\begin{tabular}{|c||c|c||c|c|c|c|c|c|c|}
			\hline\rm{model} \ref{tab: model_000_21} & \multicolumn{9}{c|}{$U(4)\times U(2)_L\times U(2)_R\times USp(6) $}\\
			\hline \hline			\rm{stack} & $N$ & $(n^1,l^1)\times(n^2,l^2)\times(n^3,l^3)$ & $n_{\Ysymm}$& $n_{\Yasymm}$ & $b$ & $b'$ & $c$ & $c'$ & 1\\
			\hline
			$a$ & 8 & $(-1,1)\times (-1,0)\times (1,1)$ & 0 & 0  & -5 & 9 & -4 & 0 & 0\\
			$b$ & 4 & $(-2,1)\times (-2,1)\times (-4,1)$ & 13 & 51  & - & - & 57 & -55 & -1\\
			$c$ & 4 & $(1,0)\times (-15,-2)\times (-1,1)$ & 13 & -13  & - & - & - & - & 0\\
			\hline
			1 & 6 & $(1, 0)\times (1, 0)\times (1, 0)$& \multicolumn{7}{c|}{$x_A = 72x_B = \frac{48}{5}x_C = 72x_D$}\\
			& & & \multicolumn{7}{c|}{$\beta^g_1=-5$}\\
			& & & \multicolumn{7}{c|}{$\chi_1=4 \sqrt{\frac{3}{5}}$, $\chi_2=6 \sqrt{15}$, $\chi_3=8 \sqrt{\frac{3}{5}}$}\\
			\hline
		\end{tabular}
	\end{center}
\end{table}

\begin{table}\scriptsize
	\caption{D6-brane configurations and intersection numbers of Model \ref{tab: model_000_22}, and its gauge coupling relation is $g^2_a=\frac{2736}{59}g^2_b=17g^2_c=\frac{85}{37}(\frac{5}{3}g^2_Y)=\frac{64}{177} 170^{3/4} \pi  e^{\phi ^4}$.}
	\label{tab: model_000_22}
	\begin{center}
		\begin{tabular}{|c||c|c||c|c|c|c|c|c|c|}
			\hline\rm{model} \ref{tab: model_000_22} & \multicolumn{9}{c|}{$U(4)\times U(2)_L\times U(2)_R\times USp(2) $}\\
			\hline \hline			\rm{stack} & $N$ & $(n^1,l^1)\times(n^2,l^2)\times(n^3,l^3)$ & $n_{\Ysymm}$& $n_{\Yasymm}$ & $b$ & $b'$ & $c$ & $c'$ & 1\\
			\hline
			$a$ & 8 & $(-1,1)\times (-1,0)\times (1,1)$ & 0 & 0  & -5 & 9 & -4 & 0 & 0\\
			$b$ & 4 & $(-2,1)\times (-2,1)\times (-4,1)$ & 13 & 51  & - & - & 63 & -65 & -1\\
			$c$ & 4 & $(1,0)\times (-17,-2)\times (-1,1)$ & 15 & -15  & - & - & - & - & 0\\
			\hline
			1 & 2 & $(1, 0)\times (1, 0)\times (1, 0)$& \multicolumn{7}{c|}{$x_A = 80x_B = \frac{160}{17}x_C = 80x_D$}\\
			& & & \multicolumn{7}{c|}{$\beta^g_1=-5$}\\
			& & & \multicolumn{7}{c|}{$\chi_1=4 \sqrt{\frac{10}{17}}$, $\chi_2=2 \sqrt{170}$, $\chi_3=8 \sqrt{\frac{10}{17}}$}\\
			\hline
		\end{tabular}
	\end{center}
\end{table}

\FloatBarrier
\section{Four-Family Standard Models from Intersecting D6-Branes with One Tilted Torus}

\begin{table}[htbp]\scriptsize
	\caption{D6-brane configurations and intersection numbers of Model \ref{tab: model_001_1}, and its gauge coupling relation is $g^2_a=g^2_b=g^2_c=(\frac{5}{3}g^2_Y)=2 \sqrt{2} \pi  e^{\phi ^4}$.}
	\label{tab: model_001_1}
	\begin{center}
		\begin{tabular}{|c||c|c||c|c|c|c|c|c|c|c|}
			\hline\rm{model} \ref{tab: model_001_1} & \multicolumn{10}{c|}{$U(4)\times U(2)_L\times U(2)_R\times USp(2)^2 $}\\
			\hline \hline			\rm{stack} & $N$ & $(n^1,l^1)\times(n^2,l^2)\times(n^3,l^3)$ & $n_{\Ysymm}$& $n_{\Yasymm}$ & $b$ & $b'$ & $c$ & $c'$ & 2 & 4\\
			\hline
			$a$ & 8 & $(1,1)\times (0,-1)\times (1,1)$ & 0 & 0  & 0 & 4 & 0 & -4 & 0 & 0\\
			$b$ & 4 & $(1,0)\times (-4,1)\times (-1,-1)$ & -3 & 3  & - & - & 0 & 0 & 0 & 4\\
			$c$ & 4 & $(0,-1)\times (4,1)\times (1,1)$ & 3 & -3  & - & - & - & - & 4 & 0\\
			\hline
			2 & 2 & $(1, 0)\times (0, 1)\times (0, -2)$& \multicolumn{8}{c|}{$x_A = 4x_B = x_C = 4x_D$}\\
			4 & 2 & $(0, 1)\times (0, 1)\times (-2, 0)$& \multicolumn{8}{c|}{$\beta^g_2=-2$, $\beta^g_4=-2$}\\
			& & & \multicolumn{8}{c|}{$\chi_1=1$, $\chi_2=4$, $\chi_3=2$}\\
			\hline
		\end{tabular}
	\end{center}
\end{table}

\begin{table}[htbp]\scriptsize
	\caption{D6-brane configurations and intersection numbers of Model \ref{tab: model_001_2}, and its gauge coupling relation is $g^2_a=2g^2_b=2g^2_c=\frac{10}{7}(\frac{5}{3}g^2_Y)=4 \pi  e^{\phi ^4}$.}
	\label{tab: model_001_2}
	\begin{center}
		\begin{tabular}{|c||c|c||c|c|c|c|c|c|}
			\hline\rm{model} \ref{tab: model_001_2} & \multicolumn{8}{c|}{$U(4)\times U(4)_L\times U(4)_R  $}\\
			\hline \hline			\rm{stack} & $N$ & $(n^1,l^1)\times(n^2,l^2)\times(n^3,l^3)$ & $n_{\Ysymm}$& $n_{\Yasymm}$ & $b$ & $b'$ & $c$ & $c'$\\
			\hline
			$a$ & 8 & $(1,-1)\times (-1,0)\times (-1,-1)$ & 0 & 0  & 0 & 4 & 0 & -4\\
			$b$ & 8 & $(0,1)\times (-1,-2)\times (1,1)$ & -2 & 2  & - & - & 0 & 0\\
			$c$ & 8 & $(1,0)\times (1,-2)\times (1,1)$ & 2 & -2  & - & - & - & -\\
			\hline
			& & & \multicolumn{6}{c|}{$x_A = \frac{1}{2}x_B = x_C = \frac{1}{2}x_D$}\\
			& & & \multicolumn{6}{c|}{}\\
			& & & \multicolumn{6}{c|}{$\chi_1=1$, $\chi_2=\frac{1}{2}$, $\chi_3=2$}\\
			\hline
		\end{tabular}
	\end{center}
\end{table}

\begin{table}[htbp]\scriptsize
	\caption{D6-brane configurations and intersection numbers of Model \ref{tab: model_001_3}, and its gauge coupling relation is $g^2_a=2g^2_b=2g^2_c=\frac{10}{7}(\frac{5}{3}g^2_Y)=4 \pi  e^{\phi ^4}$.}
	\label{tab: model_001_3}
	\begin{center}
		\begin{tabular}{|c||c|c||c|c|c|c|c|c|}
			\hline\rm{model} \ref{tab: model_001_3} & \multicolumn{8}{c|}{$U(4)\times U(4)_L\times U(4)_R  $}\\
			\hline \hline			\rm{stack} & $N$ & $(n^1,l^1)\times(n^2,l^2)\times(n^3,l^3)$ & $n_{\Ysymm}$& $n_{\Yasymm}$ & $b$ & $b'$ & $c$ & $c'$\\
			\hline
			$a$ & 8 & $(1,0)\times (-1,1)\times (-1,-1)$ & 0 & 0  & 4 & 0 & 0 & -4\\
			$b$ & 8 & $(1,-2)\times (0,-1)\times (-1,1)$ & 2 & -2  & - & - & 0 & 0\\
			$c$ & 8 & $(1,-2)\times (-1,0)\times (-1,-1)$ & 2 & -2  & - & - & - & -\\
			\hline
			& & & \multicolumn{6}{c|}{$x_A = x_B = \frac{1}{2}x_C = \frac{1}{2}x_D$}\\
			& & & \multicolumn{6}{c|}{}\\
			& & & \multicolumn{6}{c|}{$\chi_1=\frac{1}{2}$, $\chi_2=1$, $\chi_3=2$}\\
			\hline
		\end{tabular}
	\end{center}
\end{table}

\begin{table}[htbp]\scriptsize
	\caption{D6-brane configurations and intersection numbers of Model \ref{tab: model_001_4}, and its gauge coupling relation is $g^2_a=2g^2_b=g^2_c=(\frac{5}{3}g^2_Y)=\frac{8}{3} \sqrt[4]{2} \pi  e^{\phi ^4}$.}
	\label{tab: model_001_4}
	\begin{center}
		\begin{tabular}{|c||c|c||c|c|c|c|c|c|c|}
			\hline\rm{model} \ref{tab: model_001_4} & \multicolumn{9}{c|}{$U(4)\times U(4)_L\times U(2)_R\times USp(2) $}\\
			\hline \hline			\rm{stack} & $N$ & $(n^1,l^1)\times(n^2,l^2)\times(n^3,l^3)$ & $n_{\Ysymm}$& $n_{\Yasymm}$ & $b$ & $b'$ & $c$ & $c'$ & 1\\
			\hline
			$a$ & 8 & $(1,-1)\times (1,0)\times (1,1)$ & 0 & 0  & 4 & 0 & -4 & 0 & 0\\
			$b$ & 8 & $(0,1)\times (-1,2)\times (-1,1)$ & 2 & -2  & - & - & 0 & 4 & 4\\
			$c$ & 4 & $(-1,0)\times (-1,-4)\times (1,-1)$ & -3 & 3  & - & - & - & - & 0\\
			\hline
			1 & 2 & $(1, 0)\times (1, 0)\times (-2, 0)$& \multicolumn{7}{c|}{$x_A = \frac{1}{2}x_B = 2x_C = \frac{1}{2}x_D$}\\
			& & & \multicolumn{7}{c|}{$\beta^g_1=-2$}\\
			& & & \multicolumn{7}{c|}{$\chi_1=\sqrt{2}$, $\chi_2=\frac{1}{2 \sqrt{2}}$, $\chi_3=2 \sqrt{2}$}\\
			\hline
		\end{tabular}
	\end{center}
\end{table}

\begin{table}[htbp]\scriptsize
	\caption{D6-brane configurations and intersection numbers of Model \ref{tab: model_001_5}, and its gauge coupling relation is $g^2_a=g^2_b=2g^2_c=\frac{10}{7}(\frac{5}{3}g^2_Y)=\frac{8}{3} \sqrt[4]{2} \pi  e^{\phi ^4}$.}
	\label{tab: model_001_5}
	\begin{center}
		\begin{tabular}{|c||c|c||c|c|c|c|c|c|c|}
			\hline\rm{model} \ref{tab: model_001_5} & \multicolumn{9}{c|}{$U(4)\times U(2)_L\times U(4)_R\times USp(2) $}\\
			\hline \hline			\rm{stack} & $N$ & $(n^1,l^1)\times(n^2,l^2)\times(n^3,l^3)$ & $n_{\Ysymm}$& $n_{\Yasymm}$ & $b$ & $b'$ & $c$ & $c'$ & 2\\
			\hline
			$a$ & 8 & $(-1,-1)\times (0,-1)\times (-1,-1)$ & 0 & 0  & 4 & 0 & -4 & 0 & 0\\
			$b$ & 4 & $(-1,0)\times (4,1)\times (-1,1)$ & 3 & -3  & - & - & 0 & -4 & 0\\
			$c$ & 8 & $(0,-1)\times (2,-1)\times (-1,1)$ & -2 & 2  & - & - & - & - & 4\\
			\hline
			2 & 2 & $(1, 0)\times (0, 1)\times (0, -2)$& \multicolumn{7}{c|}{$x_A = 2x_B = x_C = 4x_D$}\\
			& & & \multicolumn{7}{c|}{$\beta^g_2=-2$}\\
			& & & \multicolumn{7}{c|}{$\chi_1=\sqrt{2}$, $\chi_2=2 \sqrt{2}$, $\chi_3=\sqrt{2}$}\\
			\hline
		\end{tabular}
	\end{center}
\end{table}

\begin{table}[htbp]\scriptsize
	\caption{D6-brane configurations and intersection numbers of Model \ref{tab: model_001_6}, and its gauge coupling relation is $g^2_a=2g^2_b=g^2_c=(\frac{5}{3}g^2_Y)=\frac{8}{3} \sqrt[4]{2} \pi  e^{\phi ^4}$.}
	\label{tab: model_001_6}
	\begin{center}
		\begin{tabular}{|c||c|c||c|c|c|c|c|c|c|}
			\hline\rm{model} \ref{tab: model_001_6} & \multicolumn{9}{c|}{$U(4)\times U(4)_L\times U(2)_R\times USp(2) $}\\
			\hline \hline			\rm{stack} & $N$ & $(n^1,l^1)\times(n^2,l^2)\times(n^3,l^3)$ & $n_{\Ysymm}$& $n_{\Yasymm}$ & $b$ & $b'$ & $c$ & $c'$ & 2\\
			\hline
			$a$ & 8 & $(-1,0)\times (1,1)\times (-1,1)$ & 0 & 0  & 0 & 4 & 0 & -4 & 0\\
			$b$ & 8 & $(1,2)\times (1,0)\times (1,-1)$ & -2 & 2  & - & - & 0 & -4 & 4\\
			$c$ & 4 & $(-1,4)\times (0,-1)\times (1,-1)$ & 3 & -3  & - & - & - & - & 0\\
			\hline
			2 & 2 & $(1, 0)\times (0, 1)\times (0, -2)$& \multicolumn{7}{c|}{$x_A = \frac{1}{2}x_B = \frac{1}{4}x_C = \frac{1}{4}x_D$}\\
			& & & \multicolumn{7}{c|}{$\beta^g_2=-2$}\\
			& & & \multicolumn{7}{c|}{$\chi_1=\frac{1}{2 \sqrt{2}}$, $\chi_2=\frac{1}{\sqrt{2}}$, $\chi_3=\sqrt{2}$}\\
			\hline
		\end{tabular}
	\end{center}
\end{table}

\begin{table}[htbp]\scriptsize
	\caption{D6-brane configurations and intersection numbers of Model \ref{tab: model_001_7}, and its gauge coupling relation is $g^2_a=g^2_b=2g^2_c=\frac{10}{7}(\frac{5}{3}g^2_Y)=\frac{8}{3} \sqrt[4]{2} \pi  e^{\phi ^4}$.}
	\label{tab: model_001_7}
	\begin{center}
		\begin{tabular}{|c||c|c||c|c|c|c|c|c|c|}
			\hline\rm{model} \ref{tab: model_001_7} & \multicolumn{9}{c|}{$U(4)\times U(2)_L\times U(4)_R\times USp(2) $}\\
			\hline \hline			\rm{stack} & $N$ & $(n^1,l^1)\times(n^2,l^2)\times(n^3,l^3)$ & $n_{\Ysymm}$& $n_{\Yasymm}$ & $b$ & $b'$ & $c$ & $c'$ & 1\\
			\hline
			$a$ & 8 & $(-1,0)\times (-1,-1)\times (1,-1)$ & 0 & 0  & 0 & 4 & -4 & 0 & 0\\
			$b$ & 4 & $(-1,-4)\times (-1,0)\times (1,-1)$ & -3 & 3  & - & - & 4 & 0 & 0\\
			$c$ & 8 & $(-1,-2)\times (0,1)\times (1,1)$ & -2 & 2  & - & - & - & - & 4\\
			\hline
			1 & 2 & $(1, 0)\times (1, 0)\times (-2, 0)$& \multicolumn{7}{c|}{$x_A = 2x_B = \frac{1}{2}x_C = \frac{1}{2}x_D$}\\
			& & & \multicolumn{7}{c|}{$\beta^g_1=-2$}\\
			& & & \multicolumn{7}{c|}{$\chi_1=\frac{1}{2 \sqrt{2}}$, $\chi_2=\sqrt{2}$, $\chi_3=2 \sqrt{2}$}\\
			\hline
		\end{tabular}
	\end{center}
\end{table}


\begin{table}[h]\scriptsize
	\caption{D6-brane configurations and intersection numbers of Model \ref{tb:model3}, and its gauge coupling relation is $g^2_a=\frac{8}{3}g^2_b=g^2_c=(\frac{5}{3}g^2_Y)=\frac{16}{9} 2^{3/4} \pi  e^{\phi ^4}$.}
	\label{tb:model3}
	\begin{center}
		\begin{tabular}{|c||c|c||c|c|c|c|c|c|c|}
			\hline\rm{model} \ref{tb:model3} & \multicolumn{9}{c|}{$U(4)\times U(2)_L\times U(2)_R\times USp(2) $}\\
			\hline \hline			\rm{stack} & $N$ & $(n^1,l^1)\times(n^2,l^2)\times(n^3,l^3)$ & $n_{\Ysymm}$& $n_{\Yasymm}$ & $b$ & $b'$ & $c$ & $c'$ & 3\\
			\hline
			$a$ & 8 & $(0,1)\times (1,-1)\times (1,-1)$ & 0 & 0  & 1 & 3 & -4 & 0 & 0\\
			$b$ & 4 & $(-1,1)\times (0,1)\times (-2,4)$ & 2 & -2  & - & - & -9 & -5 & 2\\
			$c$ & 4 & $(4,-1)\times (1,0)\times (1,1)$ & -3 & 3  & - & - & - & - & 0\\
			\hline
			3 & 2 & $(0, 1)\times (1, 0)\times (0, 2)$& \multicolumn{7}{c|}{$x_A = x_B = \frac{1}{2}x_C = 4x_D$}\\
			& & & \multicolumn{7}{c|}{$\beta^g_3=-4$}\\
			& & & \multicolumn{7}{c|}{$\chi_1=\sqrt{2}$, $\chi_2=2 \sqrt{2}$, $\chi_3=\frac{1}{\sqrt{2}}$}\\
			\hline
		\end{tabular}
	\end{center}
\end{table}

\begin{table}[h]\scriptsize
	\caption{D6-brane configurations and intersection numbers of Model \ref{tb:model6}, and its gauge coupling relation is $g^2_a=\frac{16}{5}g^2_b=2g^2_c=\frac{10}{7}(\frac{5}{3}g^2_Y)=\frac{16}{5} \sqrt{2} \pi  e^{\phi ^4}$.}
	\label{tb:model6}
	\begin{center}
		\begin{tabular}{|c||c|c||c|c|c|c|c|c|}
			\hline\rm{model} \ref{tb:model6} & \multicolumn{8}{c|}{$U(4)\times U(2)_L\times U(2)_R\times  $}\\
			\hline \hline			\rm{stack} & $N$ & $(n^1,l^1)\times(n^2,l^2)\times(n^3,l^3)$ & $n_{\Ysymm}$& $n_{\Yasymm}$ & $b$ & $b'$ & $c$ & $c'$\\
			\hline
			$a$ & 8 & $(0,1)\times (1,-1)\times (1,-1)$ & 0 & 0  & 1 & 3 & -4 & 0\\
			$b$ & 4 & $(-1,1)\times (0,1)\times (-2,4)$ & 2 & -2  & - & - & -6 & -6\\
			$c$ & 4 & $(2,-1)\times (1,0)\times (2,2)$ & -2 & 2  & - & - & - & -\\
			\hline
			& & & \multicolumn{6}{c|}{$x_A = x_B = \frac{1}{2}x_C = 2x_D$}\\
			& & & \multicolumn{6}{c|}{$\chi_1=1$, $\chi_2=2$, $\chi_3=1$}\\
			\hline
		\end{tabular}
	\end{center}
\end{table}

\end{document}